\documentclass[fleqn,usenatbib]{mnras}

\usepackage[T1]{fontenc}
\usepackage{ae,aecompl}

\usepackage{graphicx}	% Including figure files
\usepackage{amsmath}	% Advanced maths commands
\usepackage{amssymb}	% Extra maths symbols
\usepackage{mathrsfs}
\usepackage{hyperref}
\usepackage[dvipsnames]{xcolor}
\usepackage{newtxtext,newtxmath}

\newcommand{\LCDM}{$\Lambda$CDM }
\newcommand{\MSun}{\, \rm M_{\odot}}
\newcommand{\K}{\, \rm K}
\newcommand{\lya}{Ly$\alpha$}

\DeclareMathOperator\erf{erf}
\DeclareMathOperator\arcsinh{arcsinh}

\def\citejap#1{\citeauthor{#1}\ \citeyear{#1}}
\def\bigstrut{\vrule width0pt height1.0em}

\title[An analytic model for cosmic star formation]{Extended Hernquist-Springel formalism for cosmic star formation}

\author[D. Sorini and J. A. Peacock]{%
Daniele Sorini\thanks{E-mail: sorini@roe.ac.uk}
and John A. Peacock
\\
Institute for Astronomy, University of Edinburgh, Royal Observatory, Edinburgh, EH9 3HJ, United Kingdom\\
}

\pubyear{2021}

\begin{document}
\label{firstpage}
\pagerange{\pageref{firstpage}--\pageref{lastpage}}
\maketitle

% Abstract of the paper
\begin{abstract}
We present a revised and extended version of the analytic model for cosmic star formation originally given by Hernquist \& Springel in 2003. The key assumption of this formalism is that star formation proceeds from cold gas, at a rate that is limited by an internal consumption timescale at early times, or by the rate of generation of gas via cooling at late times. These processes are analysed as a function of the mass of dark matter haloes and integrated over the halo population. We modify this approach in two main ways to make it more general: (1) halo collapse times are included explicitly, so that the behaviour is physically reasonable at late times; (2) allowance is made for a mass-dependent baryon fraction in haloes, which incorporates feedback effects. This model reproduces the main features of the observed baryonic Tully-Fisher relationship, and is consistent with observational estimates of the baryon mass fraction in the intergalactic medium. With minimal adjustment of parameters, our approach reproduces the observed history of cosmic star formation within a factor of two over the redshift range $0<z<10$. This level of agreement is comparable to that achieved by state-of-the-art cosmological simulations.  Our simplified apparatus has pedagogical value in illuminating the results of such detailed calculations, and also serves as a means for rapid approximate exploration of non-standard cosmological models.
\end{abstract}

% Select between one and six entries from the list of approved keywords.
% Don't make up new ones.

%%%%%%%%%%%%%%%%%%%%%%%%%%%%%%%%%%%%%%%%%%%%%%%%%%

%%%%%%%%%%%%%%%%% BODY OF PAPER %%%%%%%%%%%%%%%%%%

\section{Introduction}

The existence of visible galaxies was the first and most obvious clue to the existence of a wider universe beyond the distribution of nearby stars, so understanding why galaxies exist has been a primary task of cosmological research from the very beginning. While it would be rash to claim that this problem is now solved, it is certainly true that we have a sophisticated appreciation of many of the physical mechanisms that contribute to the creation of galaxies and the formation of stars within them -- as set out in e.g. the textbook by \cite{mo_book}.

This understanding can be local, i.e. an attempt at a detailed picture of the internal structure of a galaxy that accounts for distinct bulge/disk components, spiral arms etc., or it can be global. In this latter case we focus less on galaxies as individuals and more as a single population, whose output is the overall history of star formation in the Universe. This is an interesting quantity, not least because galaxy surveys readily determine the cosmic star-formation rate density (CSFRD) in the form of $\rm M_\odot \, yr^{-1}$ per unit comoving volume. Early optically-selected deep redshift surveys showed conclusively that this quantity declined strongly from $z\approx1$ to the present, so that there was a global quenching of star formation \citep{Lilly1996}. Accounting for this cosmic shutdown of star-forming activity remains one of the principal issues in galaxy evolution. Subsequent extensions to deeper HST data and to longer wavelengths established that the CSFRD peaked at redshift $z\approx 2$, having increased by roughly an order of magnitude since $z=10$ (\citejap{Madau1996}; \citejap{Madau_rev}).

These observations present a natural challenge to theoretical models of galaxy formation. The context for this modelling is of course the standard $\Lambda$CDM background cosmology, in which the dominant process is the growth of the population of dark matter haloes via hierarchical merging. This aspect of the problem is well understood analytically (e.g. \citejap{LaceyCole1993}), and was validated by the results of large N-body cosmological simulations \citep[e.g.][]{Springel_2005, Klypin_2011, Angulo_2012, Fosalba_2015}, so the challenge is the astrophysical one of following the diffuse gas within these haloes and understanding its transmutation into stars. The contributing processes and their interactions are sufficiently complex that precise predictions require detailed hydrodynamic simulations, and decades of cumulative effort have led to the creation of a number of sophisticated codes for this purpose \citep[e.g.][]{OWLS, Almgren_2013, Enzo_Bryan2014, Dubois_2014, FIRE2014, Illustris_V2014, Lukic_2015, EAGLE_Schaye2015, Mufasa, McCarty_2017, IllustrisTNG2018, Simba_Dave2019}. 

These codes have some impressive achievements in terms of producing simulated galaxy populations with a fair degree of realism, but they are not without their difficulties. A high numerical resolution is required, meaning that large truly representative volumes are difficult to simulate. Even so, many of the physical processes of relevance remain well below the numerical resolution scales, and so have to be treated via effective `subgrid' approximations. Also, the calculations are highly demanding of computer time, so that in practice it it difficult to explore a wide range of model options. For these reasons, it is attractive to have a more rapid alternative in which the uncertainties of the subgrid processes are grafted explicitly onto a more nearly analytic treatment of the dark-matter halo population. An early and influential example of such modelling was set out by \cite{White_1991}, which then underwent subsequent refinements \citep[e.g.][]{Kauffmann1993, Cole_1994, Guideroni_1998, Kauffmann_1999, Cole2000}.  Extensions of this work encompassed the assembly of the central black hole, giving a comprehensive picture of the evolution of galaxies and quasars \citep[e.g.][]{Kauffman2000, Somerville2008, Henriques_2015, Lacey_2016}. An intermediate approach between pure Monte Carlo halo merger trees and hydrodynamical simulations is to apply the semianalytic recipes to haloes found in collisionless simulations, where the merger history is accompanied by a knowledge of the spatial distribution at any given epoch (e.g. \citejap{Croton2006}; \citejap{LGalaxies2020}).  For an overview of further literature in this field, see the review by \cite{feedback_review}.

One concern in all this work is the degree to which it is genuinely predictive. The subgrid and semianalytic recipes contain a large number of adjustable parameters, so one may be concerned that detailed models are fine tuned to $\Lambda$CDM and so would risk a lack of robustness in their predictions if cosmology were to be varied.
Such counter-factual variations are undoubtedly of interest, because cosmology contains a number of puzzling coincidences that are connected to special values of cosmological parameters. For example, there is a near equality between the present energy density of non-relativistic matter and the cosmological constant, $\Lambda$, whose value is anomalously small compared to quantum mechanical predictions. Thus the rapid decline of cosmic star formation at $z<1$ occurs just when the universe is starting to become dominated by dark energy for the first time (leaving aside a possible inflationary phase at the earliest times). Is this simply a coincidence, or could $\Lambda$ play a role in shutting down star formation, so that relatively fewer observers would be produced if $\Lambda$ were much larger? A similar coincidence exists between the era of reionization and equality between dark energy and radiation \citep{Lombriser_2017}. 
This line of thinking provokes an interest in calculating galaxy formation in non-standard cosmologies, 
and a certain amount of work of this kind has been carried out, both via direct simulation
(\citejap{Nagamine2004}; \citejap{Barnes_2018}; \citejap{Salcido_2018}) and via semianalytics (\citejap{Bousso_2010}; \citejap{sudoh2017}).

The challenge for such investigations is that it is
impractically time-consuming to explore a wide range of models
by direct simulation. Pure semianalytics are faster but have analogous limitations
in terms of limited numbers of haloes (and the need to truncate merger histories to
exclude the lowest-mass haloes entirely). One may thus consider an alternative approach, where cosmic star formation is modelled either from first principles \citep[e.g.][]{Ikea} or via nearly analytical treatment where only a few key relationships are set by empirical constraints \citep[e.g.][]{Behroozi_2013_model, Lu_2014, Moster_2018, Behroozi_2019, Grylls_2019} or inspired by numerical simulations \citep[e.g.][]{Rasera_2006, Dave_2012, Salcido_2020}.

Another example was the analytical formalism set out by \citeauthor{HS03} (\citeyear{HS03}; HS03). This influential work
offers a complementary first-principles approach that gives an
appealingly direct insight into the physical origin of its results. But some assumptions
of this framework require modification in the light of subsequent developments in
modelling galaxy formation. Most particularly, there is now an increased focus on the essential role of feedback, in which energy released in conjunction with cosmic star formation affects the progress of star formation itself (see the review by \citejap{feedback_review}); how can such processes be allowed for within the HS03 framework? The model can also benefit from modification in order to yield physically sensible results in unusual regimes, especially regarding the long-term behaviour of star formation. Given the recent shutdown of star formation, it is natural to wonder how this will continue into the future, and what the asymptotic integrated efficiency of star formation might be. But the original HS03 model fails to give sensible results in this respect, with star formation continuing indefinitely.

The aim of this paper is therefore to revisit the Hernquist-Springel approach, making allowance for feedback in the form of the removal of baryons from low-mass haloes, and correcting the treatment of star formation in the far future.
Section \S~\ref{sec:overview} gives an overview of the HS03 model, while \S~\ref{sec:formalism} presents the details of our extended formalism, emphasising the existence of two regimes of star formation, depending on whether the supply of cold gas is limited by the cooling rate. We also include here the modification of the formalism to allow for a variable halo baryon fraction, although the details are deferred to Appendix \S~\ref{sec:baryon_model}. Section \S~\ref{sec:results} then presents the predictions of the modified model, confronting with data on the cosmic star-formation history and considering its likely future behaviour. We discuss the limitations of our formalism in \S~\ref{sec:limitations}, and summarise the main conclusions of our work in \S~\ref{sec:conclusions}. Throughout, we denote units of comoving lengths with a `c' prefix (e.g. $\rm cMpc$) to distinguish them from proper length units (e.g. $\rm Mpc$).

\section{Overview of the HS03 model}
\label{sec:overview}

\subsection{Cosmic star formation rate density}
\label{sec:CSFRD}

The HS03 model computes the evolving cosmic star formation rate density (CSFRD) in a manner analogous to the halo model of clustering: as a superposition of the star formation in different haloes, integrated over the halo population. The CSFRD is thus written as
\begin{equation}
\label{eq:CSFRD}
	\dot{\rho}_{*} (z) = \bar{\rho}_0 \int g(M, \, z) \, s(M, \, z) \, d\ln M ,
\end{equation}
where $\bar{\rho}_0$ is the mean matter density of the universe (independent of $z$, since we will express the CSFRD in terms of comoving volume units), and $g(M, \, z)=dF/d\ln M$ is the halo multiplicity function, where $F(M, \, z) = 1- F_{\rm c}(M, \ z)$, with $F_{\rm c}(M, \, z)$ being the usual collapse fraction in haloes with total mass $>M$. The critical astrophysical component is
\begin{equation}
s(M, \, z) = \langle \dot{M}_* \rangle / M,
\end{equation}
which is the average normalised star formation rate (nSFR) in haloes of a given \emph{total} mass $M$.
We thus require three fundamental quantities in order to compute the CSFRD: the halo multiplicity function, the nSFR, and the lower bound of the integral in equation~\eqref{eq:CSFRD}. We will discuss $g(M, \, z)$ in this section, while we will go through the other two quantities in \S\,\ref{sec:formalism} and \S~\ref{sec:results}, respectively.

The halo multiplicity function follows entirely from the cosmological model through $F(M, z)$. We will adopt the analytic expression of $F(M, z)$ derived by HS03 from the Sheth \& Tormen formalism \citep{ST99, ST02}:
\begin{equation}
\label{eq:HMF}
	F(M, \, z) = A \left[ \erf \left( \frac{ \delta_{\rm c}}{2^{\frac{3}{4}} \sigma (M,\, z) }\right) + \frac{1}{\sqrt{2^{\frac{3}{5}} \pi}} \,\tilde{\Gamma}\left(\frac{1}{5}, \, \frac{\delta^2_{\rm c}}{2^{\frac{3}{2}} \, \sigma^2 (M,\, z)} \right) \right],
\end{equation}
where $\tilde{\Gamma}$ is the lower incomplete Gamma function, $\sigma^2(M,\,z)$ is the linear-theory fractional variance of matter density fluctuations averaged over spheres containing a mass $M$, and $A=[1+2^{-0.3} \pi^{-0.5} \Gamma(0.2)]^{-1}\approx 0.3222$ is a normalisation constant chosen such that all cosmic mass is contained in haloes. In the above equation, $\delta_{\rm c}=1.686$ is the linearly extrapolated critical density fluctuation for collapse (see e.g. \citejap{Peebles_1980}; \citejap{Percival_2005}). In principle, this linear collapse threshold is to be calculated from the spherical model, and so has a complicated dependence on the cosmological model, with $\delta_{\rm c}=1.686$ applying only for the Einstein--de Sitter universe. In practice, however, a better match to the empirical mass functions derived from cosmological simulations is obtained if $\delta_c$ is treated as a constant, independent of the cosmological model (see e.g. \citejap{Jenkins2001}; \citejap{Tinker2008}). The Sheth--Tormen expression is then close to universal, with deviations in the mass function of order 10\%, which are unimportant uncertainties in the context of the present work.

The redshift evolution of the halo mass function is encapsulated in $\sigma^2 (M,\, z)$, which is given by 
\begin{equation}
	\sigma^2(M,\,z) = D(z)^2 \int _0 ^{\infty} \frac{dk}{2 \pi^2} k^2 P(k) \left[ \frac{3j_1(kR_{\rm f})}{kR_{\rm f}}\right]^2,
\end{equation}
where $P(k)$ is the linear power spectrum of matter density fluctuations, $D$ is the linear growth factor within linear theory (normalised to unity at $z=0$), $j_1$ the first-order spherical Bessel function, and $R_{\rm f}$ the comoving spherical filter radius corresponding to a halo of mass $M=4/3 \, \pi\rho_{\rm m}(0)R_{\rm f}^3$, with $\rho_{\rm m}(0)$ being the present-time average matter density.

Following HS03, we define virial quantities in terms of the \textit{critical} density rather than the mean density. Precisely, we define the virial radius $R$ such that the typical matter density within a sphere of proper radius $R$ centred in a halo of virial mass $M$ at redshift $z$ equals a multiple $\Delta$ of the critical density $\rho_{\rm c}(z)$ at redshift $z$:
\begin{equation}
	M = \frac{4}{3}\pi \Delta \rho_{\rm c}(z) R^3 \, .
\end{equation}
We follow HS03 and adopt $\Delta=200$.
We define a characteristic virial velocity 
\begin{equation}
    V^2 = \frac{GM}{R} \, ,
\end{equation}
and we further \textit{define} the virial temperature $T$ of the halo as
\begin{equation}
\label{eq:Vvir}
	V^2 = \frac{2 k_{\rm B} T}{\mu}  \, ,
\end{equation}
where $k_{\rm B}$ the Boltzmann constant $\mu$ is the mean molecular weight. For a fully ionised plasma of primordial composition, we have $\mu \approx \,  0.6 m_{\rm p}$, where $m_{\rm p}$ is the proton mass. With equation~\eqref{eq:Vvir}, we chose to conform with the definition of the virial temperature adopted by HS03, for ease of comparison between our work and theirs.
As given, $V$ has the form of a circular velocity at the virial radius, but the conversion to a temperature in \eqref{eq:Vvir} really requires a 1D velocity dispersion. A conversion between these two velocity measures would require additional assumptions about the internal halo dynamics; but such complications would only have the effect of scaling $V$ by a dimensionless factor of order unity. We therefore follow HS03 in effectively assuming that such changes in the $T-M$ relation are unimportant in the context of 
other simplifying assumptions in the analysis.

We can now conveniently express the virial radius and virial mass of a halo of virial temperature $T$ as follows:
\begin{eqnarray}
\label{eq:Rvir}
	&&R = \sqrt{\frac{2}{\Delta}} \frac{V}{H(z)} \\
	\label{eq:Mvir}
	&&M = \sqrt{\frac{2}{\Delta}} \frac{V^3}{GH(z)} \, ,
\end{eqnarray}
where $H(z)$ is the Hubble constant at redshift $z$. As a consequence of this definition of virial quantities, the epoch-dependent Hubble parameter $H(z)$ will appear in many of the formulae in subsequent sections. 

Defining virial quantities in terms of the critical density rather than the mean density is a necessary choice in order to model cosmological star formation into the $\Lambda$-dominated far future of the Universe. If we defined the virial radius for a halo of given mass $M$ as enclosing an average density that is a factor $\Delta$ times the cosmic mean, this radius would scale as $R\propto (1+z)^{-1}$, and would diverge in the far future. But with a definition based on the critical density, the virial radius asymptotes to a finite proper length, depending on the virial mass of the halo. This is a physically sensible result: all haloes are expected to become isolated and cease accreting in the $\Lambda$-dominated future, so that their proper size should freeze out.

In the next subsection, we will focus on the normalised SFR, $s(M,z)$. However, we will first make one significant alteration to the formalism, using equations \eqref{eq:Vvir} and \eqref{eq:Mvir} to transform from mass as the fundamental variable that describes the halo population, to virial temperature:
z\begin{equation}
    T = \frac{\mu}{2 k_{\rm B}} \left[ \sqrt{\frac{\Delta}{2}} G H(z_{\rm c}) M \right]^{\frac{2}{3}},
\end{equation}
where $z_c$ is the collapse redshift, which in the Press-Schechter (and Sheth \& Tormen) view would be taken as the redshift under study \citep{PS74}. The advantage of working in terms
of temperature are twofold. Physically, a major determining factor
for cosmic star formation is the supply of cold gas via radiative cooling, which has a strong direct dependence on temperature. Secondly, halo virial temperature evolves more slowly than mass. The
characteristic cutoff in the mass function is given when the rms fractional density fluctuation as a function of scale is of order unity, and this rms scales as $M^{-b}$, where $b=(3+n_{\rm eff})/6$ in terms of the effective slope of the power spectrum. During the matter-dominated era, this rms scales as $a(t)$, and thus the characteristic mass scales as $a^{1/b}$ -- so that the characteristic virial temperature scales as $a^{-1+2/b}$.
For an effective spectral index of $-1$, there is thus no evolution in temperature. CDM-family power spectra tend to be more negative
than this on galaxy scales, but even so the evolution is slow. Thus working with temperature as the fundamental variable includes to a good extent the operation of the merging hierarchy, which boosts halo masses to ever higher values even as their virial temperatures remain approximately constant.

\subsection{Regimes of star formation in a single halo}

With the halo multiplicity function in place, our strategy for determining the CSFRD, as set out in \S\,\ref{sec:CSFRD}, now requires us to model the nSFR in a single halo. This section gives an overview of our approach.

To begin with, we assume that the baryonic matter within a halo consists of gas and stars only, i.e. we do not consider other components such as dust. Within this simplified view, the SFR is directly related to the gas mass $M_{\rm gas}$ in the halo, and the rate at which gas undergoes radiative cooling and is subsequently converted into stars. Thus, the SFR will be determined by whichever process is slower: the cooling time scale or the gas consumption time scale. There will thus be two different regimes, each defined by the physical process that acts as the bottleneck for star formation. As we will discuss in \S\,\ref{sec:time-scales}, cooling is rapid and efficient at high redshift, so that the SFR is set by the gas consumption time scale, and the resulting SFR is simply proportional to the gas mass in the halo (see \S\,\ref{sec:high-z}). On the other hand, at low redshift the SFR is cooling-driven. In \S\,\ref{sec:low-z} we will show that adopting a spherically symmetric power-law gas density profile yields an analytic expression for the SFR, which can be easily expressed as a function of $M_{\rm gas}$.

The available gas mass for star formation is of course affected by feedback processes such as stellar winds, energy from supernova explosions, or jets ejected by active galactic nuclei (AGN). A major effect of these processes is to alter the baryon content of haloes, and we have therefore developed a model to determine the baryonic mass fraction of haloes as a function of their virial temperature and redshift. This constitutes an improvement with respect to the HS03 formalism, whereby the baryonic mass fraction of all haloes was implicitly assumed to match the Universal baryon fraction $f_{\rm b}=\Omega_{\rm b}/\Omega_{\rm m}$. The details are given in Appendix \ref{sec:baryon_model}, where we also consider previous relevant work on the subject by \cite{Rasera_2006}. As we will discuss in \S~\ref{sec:baryons}, even though our model only includes the effect of supernovae and not AGN, it broadly reproduces the observed correlation between baryonic mass fraction and rotational velocity of galaxies \citep{Lelli_2016} known as baryonic Tully-Fisher relation (bTFR; \citealt{bTFR}), as well as the index of the Kennicutt-Schmidt relationship measured by \cite{Kennicutt_1998}. It also provides reasonable predictions for the mass fraction of baryons locked in the intergalactic medium (IGM). 
In conclusion, our strategy will be to assume our own model for the baryonic mass fraction of haloes and to compute the nSFR in the high-$z$ and low-$z$ regimes mentioned earlier. We will then define a global nSFR by connecting the two solutions with a sufficiently generic interpolating function. The resulting nSFR will then provide $s(M,\,z)$ in equation~\eqref{eq:CSFRD}, which will give a prediction of the CSFRD. The details of the formalism are presented in the next section.

\section{Formalism}
\label{sec:formalism}

\subsection{Time scales regulating star formation}
\label{sec:time-scales}

Star formation is affected by the complex interplay of gas cooling and feedback processes such as supernovae-driven winds or AGN jets from central black holes. Within the current simplified approach we cannot expect to  model such feedback mechanisms in detail, but we can still implicitly account for them in our reasoning for the estimation of the SFR. Indeed, the key point of our model is that star formation is regulated by two fundamental time scales: the cooling time, and the gas consumption time scale. The question is how such time scales are affected by feedback, and which one dominates star formation at different cosmic times.

At low enough redshifts, lower characteristic densities mean that cooling will be slow relative to the gas consumption timescale, meaning that new cold gas is processed into stars as soon as it is generated. The process then becomes supply-limited, so that the SFR for a given halo is expected to be proportional to the gas cooling rate. To keep our analytic treatment feasible, we adopt the simplifying assumption that the dynamical equilibrium between star formation, cooling and feedback responds linearly to variations in the cooling rate (see \S~3.4 in HS03). In this idealised view, the cooling rate can still be seen as a proxy for the SFR. We thus distinguish two different regimes of star formation, depending on the dominant time scale at the redshift considered. Following HS03, we will describe the cooling-limited SFR at low redshift in \S\,\ref{sec:low-z}, and the gas-consumption-limited star formation regime in \S\,\ref{sec:high-z}. 

\subsection{Low-redshift regime}
\label{sec:low-z}

In the low-redshift cooling-dominated regime of star formation, HS03 estimated the production of cold gas by using the concept of a cooling front.
Because cooling is more efficient at higher density, the gas in the innermost regions of the halo will cool first, followed by shells at progressively larger radii. We can thus visualise the cooling process as the expansion of a cooling front from the core of the halo outwards. At any given time, we can then define a cooling radius, within which gas has cooled and remains cool thereafter, subsequently undergoing star formation. The local cooling time sets the extent of the cooling radius.

It must be admitted that this description is heavily idealised and only includes part of the demographics of cold gas in the Universe. We know this to be so on observational grounds, since the total comoving cosmic density of neutral hydrogen has apparently not changed since $z=2$, even though star formation has declined precipitously over that period \citep[e.g.][]{Lanzetta_1991, Prochaska2009}. The answer to this apparent disparity must be that the total HI measurements, which derive from damped Ly$\alpha$ systems, are dominated by rather diffuse gas which is not in practice a reservoir for star formation. This situation was well known at the time HS03 was written, and we will follow their simplifying assumption that a cooling-front model in haloes is capable of accounting for the generation of the denser cold gas that is relevant for star formation.

The evolution of the cooling front will depend in detail on the density profile of the halo. It is well known that DM density profiles found in N-body cosmological simulations can be universally described by an NFW profile (\citejap{NFW}). This profile can be locally approximated with a power law, and we follow HS03 in adopting the stronger simplifying approximation that both the DM and gas density profiles are spherically symmetric and described by a power law over the full extent of the halo. As such, for the gas density we have
\begin{equation}
\label{eq:dens_prof}
	\rho _{\rm gas} (r) = (3-\eta) \frac{M_{\rm gas}}{4 \pi R^{3} } \left(\frac{R}{r} \right)^{\eta}  \, ,
\end{equation} 
where the slope $\eta > 0$ is a free parameter of the model. This power-law density profile requires truncation at the virial radius, so we consider only gas within $R$ to be part of the halo, and indeed the volume integral of equation~\eqref{eq:dens_prof} out to $r=R$ is equal to $M_{\rm gas}$. The total density profile follows an analogous profile, where $M_{\rm gas}$ needs to be replaced with $M(z)$. While we do not need to worry about the DM distribution for the determination of the nSFR, this will constitute an important assumption for our modelling of the baryon mass fraction in haloes (see \S\,\ref{sec:baryons} and appendix \S\,\ref{sec:baryon_model}). 

The effective power-law slope, $\eta$, can be chosen based on results of simulations, or from data-driven considerations. There are however some physical constraints on $\eta$: we must have $\eta<3$, otherwise equation~\eqref{eq:dens_prof} would imply an infinite $M_{\rm gas}$. Furthermore, we cannot have $\eta\leq 0$, since the halo density will always fall with radius. For $\eta=2$, one would have an isothermal profile; this is the slope at the characteristic radius of the NFW profile, $r_s$. HS03 set $\eta = 1.65$, which they found to be the best fit to the results of the simulations in \cite{SH03_sims}. We will later set it to $15/7 \approx 2.14$, as this value best matches observables such as the index of the Kennicutt-Schmidt law \citep{Kennicutt_1998} and the bTFR measured by \cite{Lelli_2016} (see \S\,\ref{sec:baryons} and appendix~\ref{sec:baryon_model}).

The local cooling time at distance $r$ from the centre of the halo is
\begin{equation}
\label{eq:tcool}
	t_{\rm cool} = \frac{3 k_{\rm B} T \rho_{\rm gas} (r)}{2 \mu n_{\rm H}(r)^2 \Lambda (T)} \, ,
\end{equation}
where $n_{\rm H}$ is the number density of hydrogen, and $\Lambda(T)$ is the cooling function. Adopting the same assumption as in HS03, we consider a primordial cooling function \citep{Sutherland_1993}, meaning that we are ignoring the effect of metal cooling. We will discuss the impact of this assumption in \S\,\ref{sec:results}.

Following the cooling front argument presented earlier, at any time $t$ a gas mass \smash{$M_{\rm cool}$} cools out to the cooling radius $r_{\rm cool}(t)$, and remains cool thereafter. We can thus insert the expression of the gas density in equation~\eqref{eq:dens_prof} into equation~\eqref{eq:tcool}, and easily obtain the cooling radius if the cooling time of interest, $t_{\rm cool}$, is known at time $t$:
\begin{equation}
\label{eq:rcool}
    \frac{r_{\rm cool}(t)}{R}  = \left[ \frac{(3-\eta) \mu X^2 M_{\rm gas} \Lambda (T)}{6 \pi k_{\rm B} T m_{\rm H}^2 R^3} t_{\rm cool}(t) \right]^{\frac{1}{\eta}} \, ,
\end{equation}
with $X=0.76$ and $m_{\rm H} \approx m_{\rm p}$ being the cosmic mass fraction of hydrogen and its atomic mass, respectively. Assuming that the gas density profile does not evolve appreciably during cooling, we can therefore write
\begin{equation}
\label{eq:cool_rate}
	\frac{d M_{\rm cool} }{d t} = 4 \pi \rho _{\rm gas} (r_{\rm cool}) r^2_{\rm cool} \frac{d r_{\rm cool}}{dt} \, ,
\end{equation}
and with $r_{\rm cool}(t)$ given by equation~\eqref{eq:rcool}, we obtain:
\begin{equation}
\label{eq:cool_rate_gen}
    \frac{d M_{\rm cool} }{d t} = S(T) \left( \frac{M_{\rm gas}(T, \,z)}{f_{\rm b} M_0(T)} \right)^{\frac{3}{\eta}} \left(\frac{H(z)}{H_0} \right)^{\frac{9}{\eta}-3} (H_0 t_{\rm cool})^{\frac{3}{\eta}-2} \frac{dt_{\rm cool}}{dt} \, .
\end{equation}
In the above equation, $M_0(T)$ is the total virial mass of the halo at redshift $z=0$, and $H_0$ is the $z=0$ Hubble constant. We also conveniently defined the temperature-dependent quantity $S(T)$, which has the dimensions of mass over time. In this way, if we consider the cooling rate as a proxy for the SFR, $S(T)$ represents the SFR of a halo with virial temperature $T$ and gas mass $M_{\rm gas}=f_{\rm b} M_0(T)$ at redshift $z=0$. Using the definitions~\eqref{eq:Vvir}-\eqref{eq:Rvir}, we can explicitly determine $S(T)$ by expressing all virial quantities that enter equation~\eqref{eq:cool_rate_gen} via equation~\eqref{eq:rcool} in terms of the virial temperature of the halo. We obtain
\begin{equation}
    S(T) = C(T,\,\eta) \frac{ m_{\rm H}^2}{ X^2 H_0 \Lambda(T)} \left(\frac{2k_{\rm B}T}{\mu} \right)^{\frac{5}{2}}  , \, 
\end{equation}
where $C(T,\, \eta)$ is now a dimensionless constant defined by
\begin{equation}
    C(T,\, \eta) = \frac{3\pi}{\eta} \left(\frac{2}{\Delta}\right)^{\frac{3}{2}} \left[ \frac{\Delta}{2} \frac{(3-\eta) f_{\rm b} X^2 \mu H_0 \Lambda(T)}{6 \pi G m_{\rm H}^2  k_{\rm B} T} \right]^{\frac{3}{\eta}} \, .
\end{equation}

To summarise, equation~\eqref{eq:cool_rate} tells us that the fraction of cool gas increases due to expansion of the cooling front. By definition, the extent of the cooling front at any time $t$ is set by $t_{\rm cool}(r_{\rm cool}(t))=t$, in which case it seems that we can  immediately use equation~\eqref{eq:cool_rate_gen} to 
determine the time evolution of the gas cooling rate, and hence of the SFR. But before doing so, we must think carefully about the physical meaning of the independent time variable $t$. 

In principle, $t$ should be the time since the halo was formed, i.e. the time since collapse or since the last major merger \citep{Somerville_1999}. In this view, the time $t$ in general depends on the mass of the halo considered, and on the cosmological era. In the matter-dominated era, low-mass haloes survive for a time comparable to the age of the Universe at the redshift of interest, while massive haloes beyond the exponential cutoff of the halo mass function live for a shorter time. It is important to stress that we are actually interested in determining the amount of cool gas in the average halo, that is to say we need to consider an ensemble average of haloes with virial temperature $T$. Therefore, because massive haloes are rare, one can reasonably argue that $t$, and hence $t_{\rm cool}$, should be of the order of the age of the Universe in the matter-dominated era \citep[this argument follows][]{White_1991}. On the other hand, \cite{Springel_2001} argued that $t_{\rm cool}$ should be approximately equal to the dynamical time of the halo $t_{\rm dyn} \equiv R/V = (2/\Delta)^{1/2} H(z)^{-1}$, as it is on this time scale that the gas profile reacts to pressure losses from cooling, and consequently the dynamical time should set the extent of the cooling radius. For this reason, HS03 set $t_{\rm cool}=t_{\rm dyn}$ and solved equation~\eqref{eq:cool_rate_gen} accordingly.

While this choice was shown to yield good agreement with simulations \citep{Yoshida_2002}, its validity breaks down in the far $\Lambda$-dominated future. Once merging ceases, haloes will exist in isolation and will be able to cool without interruption for an unlimited timespan. Thus, the cooling time will eventually become much larger than the dynamical time, which instead asymptotes to $(2 /\Delta \Omega_{\Lambda})^{1/2} H_0^{-1}$ as cosmic time tends to infinity. 
In our formalism, we will therefore assume that the cooling time equals the cosmic time $t$ far enough in the future, while converging to a multiple $f_{\rm dyn}$ of the dynamical time $t_{\rm dyn}$ at early cosmic times. Specifically, we set
\begin{equation}
\label{eq:tcool_smooth}
    t_{\rm cool} (t) = f_{\rm dyn} t_{\rm dyn} \left[ 1 -E + \left(\frac{t}{f_{\rm dyn} t_{\rm dyn}}\right)^m \right]^{\frac{1}{m}} \, ,
\end{equation}
where $m$ is a softening parameter that allows for a smooth transition between the early and late times regimes. The counterterm, $E$, is needed in order to force the desired early-time asymptote of
$t_{\rm cool} = f_{\rm dyn} t_{\rm dyn}$, because $t/t_{\rm dyn}$ tends to a constant in the early matter-dominated phase rather than vanishing at early times. Thus the required counterterm is
\begin{equation}
    E = \left(\frac{2}{3 f_{\rm dyn}} \sqrt{\frac{\Delta}{2}} \right)^m \, .
\end{equation}
Recalling the analytic solution for the time-dependence of the scale factor in a flat \LCDM universe,
\begin{equation}
    a(t) = \left(\frac{\Omega_{\rm m}}{\Omega_{\Lambda}}\right)^{\frac{1}{3}} \left[ \sinh \left(\frac{3}{2} \sqrt{\Omega_{\Lambda}} H_0 t \right) \right]^{\frac{2}{3}} \, ,
\end{equation}
we can re-cast equation~\eqref{eq:tcool_smooth} in terms of redshift:
\begin{equation}
\label{eq:tcool_final}
    t_{\rm cool}(z) = \sqrt{\frac{2}{\Delta}} \frac{f_{\rm dyn}}{H(z)} \left[  1 -E +  A(z)^m \right]^{\frac{1}{m}} \, ,
\end{equation}
where
\begin{equation}
    A(z) = \frac{2}{3 f_{\rm dyn}} \sqrt{\frac{\Delta}{2\, \Omega_{\Lambda}}} \frac{H(z)}{H_0} \arcsinh \left(\sqrt{\frac{\Omega_{\Lambda}}{\Omega_{\rm m} (1+z)^3}} \right) \, .
\end{equation}

\begin{figure}
	\includegraphics[width=\columnwidth]{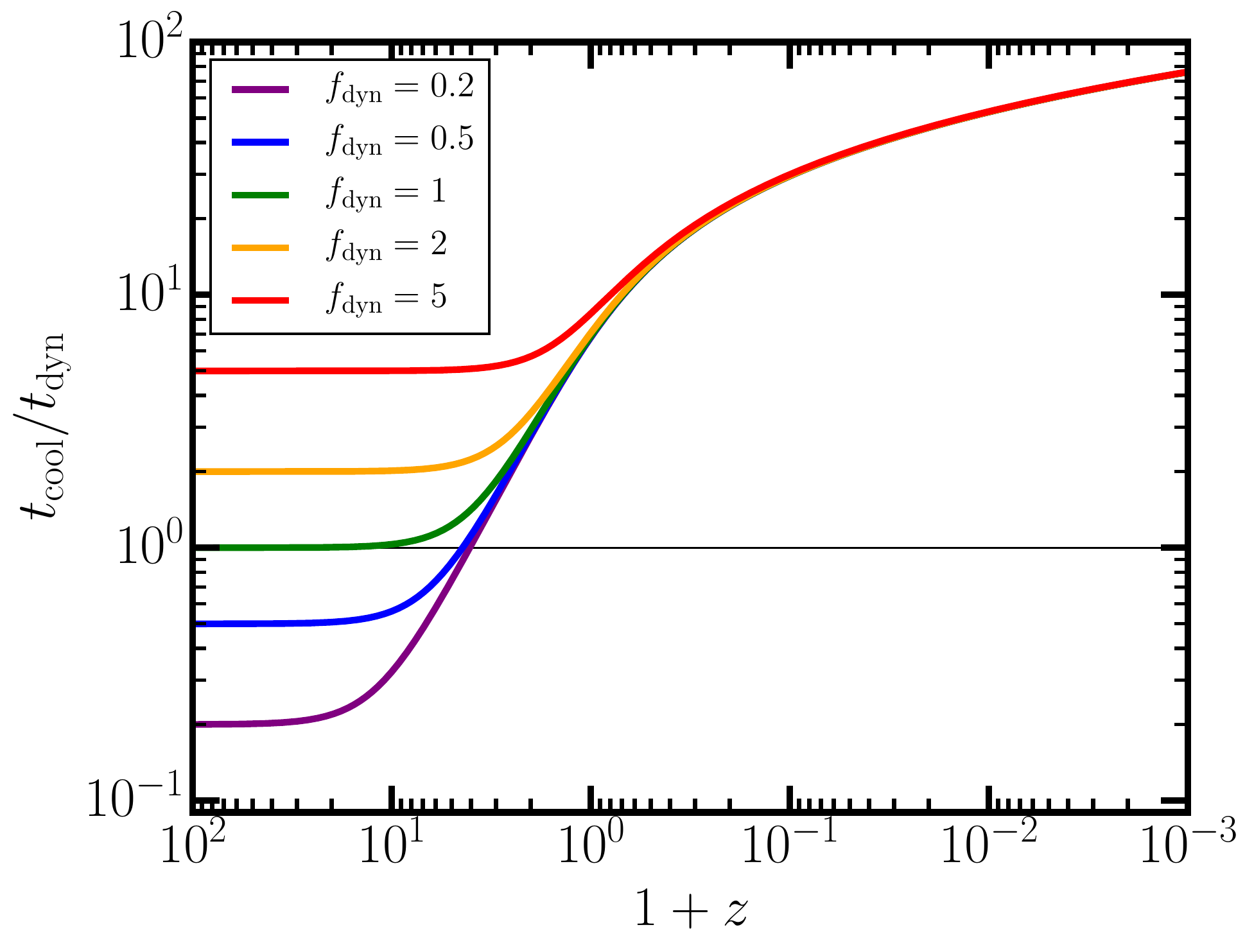}
    \caption{Ratio between the cooling time and dynamical time of haloes as a function of redshift, computed according to equation~\eqref{eq:tcool_final}. By construction, each line converges to a multiple $f_{\rm dyn}$ of the dynamical time at high redshift: $f_{\rm dyn}=0.2$, $f_{\rm dyn}=0.5$, $f_{\rm dyn}=1$, $f_{\rm dyn}=2$ and $f_{\rm dyn}=5$ for the purple, blue, green, orange and red lines, respectively. At negative redshift, i.e. in the future, the cooling time asymptotes to the age of the universe irrespective of $f_{\rm dyn}$. In all cases, the smoothing parameter in equation~\eqref{eq:tcool_final} is set to $m=2$.} 
\label{fig:tcool}
\end{figure}

We show that equation~\eqref{eq:tcool_final} yields the desired behaviour for $t_{\rm cool}(z)$ in Figure~\ref{fig:tcool}, where we plot the ratio $t_{\rm cool}/t_{\rm dyn}$ as a function of redshift. Each line corresponds to a different value of $f_{\rm dyn}$, and follows the colour coding specified in the legend of the plot. We also added  a horizontal black line corresponding to $t_{\rm cool}=t_{\rm dyn}$ to guide the eye. At late times (i.e., negative redshift), all lines converge to the same solution, as $t_{\rm cool}$ equals the age of the Universe in the $\Lambda$-dominated era. Conversely, at early times (i.e., high redshift), the lines asymptote to different values, according to the corresponding $f_{\rm dyn}$. 
In plotting Figure~\ref{fig:tcool}, we chose $m=2$ for the smoothing parameter in equation~\eqref{eq:tcool_final}, and this will be our default value. The presentation below will keep $m$ unspecified;
the final results for the CSFRD are only marginally affected by its specific numerical value.

At this point, we can finally determine the redshift dependence of the SFR for an ensemble of haloes with a given virial temperature $T$. To do that, we need to insert the expression of $t_{\rm cool}(z)$ given by equation~\eqref{eq:tcool_final} in the right hand side of equation~\eqref{eq:cool_rate_gen}, where we also need to re-cast the derivative of $t_{\rm cool}$ with respect to redshift rather than time. As explained in \S\,\ref{sec:time-scales}, at low redshift the bottleneck for star formation is represented by the gas cooling rate. In this regime, we can thus identify the SFR with the gas cooling rate.
After some manipulation (verified with \textit{Mathematica}), we obtain the cooling-driven SFR ($\dot{M}_*=\dot{M}_{\rm cool}$):
\begin{multline}
\label{eq:SFR}
      \dot{M}_*(z) = S(T) \left( \frac{H(z)}{H_0} \right)^{\frac{6}{\eta}-1} \left(\frac{M_{\rm gas}}{f_{\rm b}M_0(T)} \right)^{\frac{3}{\eta}} \\
    \left(\sqrt{\frac{2}{\Delta}} f_{\rm dyn} \right)^{\frac{3}{\eta}-2} \left( 1- E + A(z)^m \right)^{\frac{3-\eta}{m \eta}-1} \\
    \left[A(z)^{m-1} + (1-E)  \frac{3 f_{\rm dyn}}{2} \sqrt{\frac{2}{\Delta}} \left(\frac{H_0}{H(z)}\right)^2 \Omega_{\rm m} (1+z)^3 \right]  \, .
\end{multline}
To compute the normalised SFR at low redshift, i.e. $s_{\rm low}(T,\, z) = \dot{M}_*/M$, we can simply divide both sides of equation~\eqref{eq:SFR} by the virial mass of a halo with fixed virial temperature $T$. This yields
\begin{multline}
\label{eq:nSFR}
      s_{\rm low}(T,\,z) = \tilde{S}(T) \left( \frac{H(z)}{H_0} \frac{f_{\rm gas}(T, \, z)}{f_{\rm b}} \right)^{\frac{3}{\eta}} 
    \left( 1- E + A(z)^m \right)^{\frac{3-\eta}{m \eta}-1} \\
    \left[A(z)^{m-1} + (1-E)  \frac{3 f_{\rm dyn}}{2} \sqrt{\frac{2}{\Delta}} \left(\frac{H_0}{H(z)}\right)^2 \Omega_{\rm m} (1+z)^3 \right]  \, ,
\end{multline}
where we now define $M_{\rm gas} = f_{\rm gas}(T,\, z) M(T, \,z)$, so that $f_{\rm gas}(T,\, z) $ is the gas mass fraction in a halo with virial temperature $T$ at redshift $z$. Similarly to equation~\eqref{eq:cool_rate_gen}, we now define the temperature-dependent quantity $\tilde{S}(T)$, with dimensions of $\rm time^{-1}$. It therefore represents the normalised SFR of a hypothetical halo with virial temperature $T$ and a gas fraction equal to the cosmic baryon mass fraction at $z=0$. The explicit expression for this quantity is
\begin{equation}
\label{eq:SFR_final}
    \tilde{S}(T) =  \frac{1}{\eta} \left[ \sqrt{\frac{\Delta}{2}} \frac{(3-\eta) f_{\rm dyn} f_{\rm b} X^2 \mu H_0 \Lambda(T)}{6 \pi G m_{\rm H}^2  k_{\rm B} T} \right]^{\frac{3}{\eta}} \frac{6 \pi G m_{\rm H}^2 k_{\rm B} T}{ f_{\rm dyn}^2 X^2 \mu \Lambda(T)} \, .
\end{equation}

The final ingredient needed for the computation of the nSFR is $f_{\rm gas}$, which is also expected to depend on the virial temperature of the halo and on redshift. To keep our treatment as analytical as possible, we adopt the approximation $f_{\rm gas}\approx f_{\rm b,\,halo}(T,\, z)$, where $f_{\rm b, \, halo}$ is the baryon  mass fraction of the halo; the expression for this will be discussed in \S~\ref{sec:baryons} and thoroughly derived in Appendix \S~\ref{sec:baryon_model}. This approximation is expected to be good at high redshift, although less exact at lower redshift if star formation has been efficient. 
Nevertheless, we did explore the implications of dropping this assumption. The first complication manifests itself in the need of coupling equation~\eqref{eq:SFR} with the mass conservation equation $M_{\rm b} = M_* + M_{\rm gas}$. Differentiating with respect to time, and then re-casting it in terms of redshift rather than time, we end up with a differential equation in $M_{\rm gas}$, which needs to be solved numerically. Carrying out this step makes the code much slower, without yielding an appreciable improvement of the match between predicted and observed CSFRD. For this reason, and also to adhere to the original philosophy of the approach, we prefer to make the simpler assumption and thus keep the model nearly analytical.

Before moving to the high-redshift regime in the next subsection, we note that if we assign the cosmic baryon fraction $f_{\rm gas} = f_{\rm b}$ to all haloes, and with the replacements $E, \, A(z) \rightarrow 1$, then equation~\eqref{eq:tcool_final} becomes $t_{\rm cool}(z) = f_{\rm dyn} t_{\rm dyn}$ and, for $f_{\rm dyn}=1$, equation~\eqref{eq:nSFR} reduces to the one found by HS03. This clearly shows that our expression for the low-redshift nSFR contains the HS03 result as a special case. 

\subsection{High-redshift regime}
\label{sec:high-z}

As discussed in \S\,\ref{sec:time-scales}, we know that at high redshift the bottleneck for star formation is represented by the gas consumption time scale. Likewise, expressing all virial quantities in the right hand side of equation~\eqref{eq:rcool} in terms of the virial temperature (which is fixed in our formalism), and considering that at early times $t_{\rm cool} \sim t_{\rm dyn}$, we can see that for $z\gg 1$ 
\begin{equation}
    \frac{r_{\rm cool}}{R} \sim \left[\frac{(3-\eta) X^2 f_{\rm b}}{6\pi} \left(\frac{\Delta}{2}\right)^{\frac{1}{2}} \frac{\mu \Lambda(T)}{k_{\rm B} T m_{\rm H}^2 G} H(z) \right]^{\frac{1}{\eta}} \, .
\end{equation}
The redshift-independent factor that multiplies $H(z)$ in the right-hand side of the equation above has the dimensions of a time. As a reference, for haloes with virial temperature $T=10^6 \, \rm K$ and for $\eta=2$, this timescale corresponds to $\sim 0.28 \, \rm Gyr$. Because $H(z)$ increases with redshift, so does $r_{\rm cool}/R$: at sufficiently early times the cooling radius will therefore reach the virial radius and all gas within the halo will be available to form stars. 

In this regime we should therefore focus on the average gas consumption time scale, $\langle t_* \rangle$. A first guess might set this locally to be of the order of the free-fall time, \smash{$t_{\rm ff}(r) \propto \rho_{\rm gas}(r)^{-1/2}$}; but haloes have an internal density that is a multiple of the mean density, so this would scale in proportion to the age of the Universe, leading to a divergent $\mathrm{SFR} = M_{\rm gas}/\langle t_* \rangle$.
For a more realistic treatment of the problem, we can look to the multiphase model for star formation that was implemented in cosmological SPH simulations by \cite{SH03_sims, SH03_SFR}. This amounts to a subgrid treatment that includes elements of self-regulation from supernova feedback. The conclusion from this work regarding $\langle t_* \rangle$ was very different: this parameter showed no dependence on cosmological epoch, nor indeed on the virial temperature of the halo under study. This permits a massive simplification of the problem, in which $\langle t_* \rangle$ is treated as a single free parameter governing the high-redshift regime. We shall follow HS03 and make the same assumption.

We thus parametrise the nSFR at high redshift following HS03:
\begin{equation}
\label{eq:M*_high-z}
   s_{\rm high}(T, \, z) = \frac{(1-\beta) x f_{\rm gas } (T,\, z) }{\langle t^* \rangle} \, ,
\end{equation}
where  $x$ is the fraction of cold gas clouds and $\beta$ is the  the mass fraction of massive ($> 8 M_{\odot}$) short-lived stars. Referring to their multiphase model for star formation \citep{SH03_sims, SH03_SFR}, HS03 set $x=0.95$, corresponding to the fraction of cold clouds at the gas density threshold above which star formation occurs. They also assume a \cite{Salpeter_1955} initial mass function (IMF) with slope $-1.35$ and upper lower limits of 40 and 0.1 $M_{\odot}$, respectively, obtaining $\beta=0.1$. We prefer to adopt the more recent \cite{Chabrier_2003} IMF, with a lower limit of 0.1 $M_{\odot}$ and a high-mass cutoff of $100\, \rm M_{\odot}$ to reflect the higher stellar masses observed in the Arches cluster (\citealt{Blum_2002, Figer_2002}; see also \citealt{Figer_2005}). This yields $\beta=0.21$, although the exact value of this parameter is not critical: as we will show later in this section, the ratio $(1-\beta)/\langle t_* \rangle$ in equation~\eqref{eq:M*_high-z} is constrained by observations, and is in effect a single free parameter. However, the mass fraction of massive stars will affect the estimation of the baryon mass fraction in haloes (see appendix~\ref{sec:baryon_model}), so there is still merit in choosing a value of $\beta$ that is consistent with observational constraints.

At high redshift few stars formed, so that the gas mass fraction within haloes can reasonably be approximated with the baryon mass fraction, i.e. $ f_{\rm gas } (T,\, z) \approx  f_{\rm b,\,halo } (T,\, z)$. We stress that this assumption is much milder than the HS03 approximation that $f_{\rm b,\,halo } (T,\, z)$ is equal to the cosmic baryon mass fraction $f_{\rm b}$. We will discuss the explicit form of $f_{\rm b,\,halo } (T,\, z)$ in Appendix~\ref{sec:baryon_model}.

The last variable that we need to determine in equation~\eqref{eq:M*_high-z} is the gas consumption time scale. This time scale was determined by HS03 from the results of hydrodynamic simulations \citep{SH03_sims}, such that the simulated galaxies reproduced the observed Kennicutt-Schmidt relation \citep{Kennicutt_1998} at $z=0$. While several cosmological simulations have succeeded in reproducing a plethora of observations related to star formation (e.g., the star formation efficiency \citealt{Guo_2011, Moster_2013, Behroozi_2013}, the evolution of the star formation rate density \citealt{Behroozi_2013, Oesch_2015}, the black-hole-stellar-mass relationship within galaxies \citealt{Kormendy_2013, McConnell_2013}, the gas fraction within haloes \citealt{Giodini_2009, Lovisari_2015}, the stellar mass function \citealt{Baldry_2008, Baldry_2012, Bernardi_2013, D_Souza_2015}, and the stellar half-mass radii of galaxies \citealt{Baldry_2012, Shen_2003}), they often do so with significantly different prescriptions for physical processes occurring on galactic and sub-galactic scales (for a review of feedback prescriptions, see \citejap{feedback_review}). For this reason, we prefer to determine the physical parameters of our model directly from observational constraints, rather than simulations.

The Kennicutt-Schmidt relation is set by the physical processes regulating star formation within haloes, which are decoupled from the Hubble flow, and we will therefore assume that the Kennicutt-Schmidt relation holds for star-forming haloes at any redshift. This assumption is backed by the \cite{Genzel_2010} observations of star-forming galaxies in the redshift range $ 0 \lesssim z\lesssim 2.3$ and sub-mm galaxies in the redshift range $1 \lesssim z \lesssim 3.5$. The observed correlation between the surface density of star formation $\Sigma_{\rm SFR}$ and the surface density of molecular gas $\Sigma_{\rm mol \, gas}$ takes a power-law form:
\begin{equation}
\label{eq:Genzel_2010}
    \frac{\Sigma_{\rm SFR}}{\rm M_{\odot} yr^{-1} kpc^{-2}} = 10^A \left( \frac{\Sigma_{\rm mol\, gas}}{\rm M_{\odot} pc^{-2}} \right)^B \, ;
\end{equation}
the best-fit parameters are $A=-3.48\pm 0.21$ and $B=1.17 \pm 0.09$ (where these published error bars are formally $3\sigma$).

In our formalism, we assume $\Sigma_{\rm SFR}=\mathrm{SFR}/\pi R^2$, and using equation~\eqref{eq:M*_high-z} we have
\begin{equation}
\label{eq:KS}
    \Sigma_{\rm SFR} = \frac{1-\beta}{\langle t^* \rangle} \frac{x M_{\rm gas}}{\pi R^2} =  \frac{1-\beta}{\langle t^* \rangle} \, \Sigma_{\rm mol\, gas} \, ,
\end{equation}
where in the last equality we used $M_{\rm mol\, gas} = x M_{\rm gas}$, because molecular gas generally resides in cold clouds \citep[see the review by][]{mol_clouds_rev}.
Comparing equations~\eqref{eq:KS} and~\eqref{eq:Genzel_2010}, we see that in our case $B=1$, which is slightly lower than the value given by \cite{Genzel_2010}. But that paper clearly felt that the uncertainties on the slope could be subject to systematics, so the formal discrepancy does not seem a cause for concern.

Requiring our normalisation factor $(1-\beta)/\langle t^* \rangle$ to match the data in \cite{Genzel_2010}, we obtain $\langle t^* \rangle = 2.39 \, \rm Gyr$. This is almost a factor of two larger than the average gas consumption scale adopted by HS03 $\langle t^* \rangle= 1.4 \, \rm Gyr$. Nevertheless, both this choice of $\langle t^* \rangle$ and ours fall within the range $0.5-3 \, \rm Gyr$ that is observationally inferred for typical star-forming galaxies with gas surface densities between $10 \MSun \, pc^{-2}$ and $1000 \MSun \, pc^{-2}$ \citep{Kennicutt_Evans_2012}.

Because $\langle t_* \rangle$ is a constant and the dynamical time of a halo goes as $H(z)^{-1}$, the ratio $\langle t_* \rangle / t_{\rm dyn}$ increases indefinitely at higher redshift, while it asymptotes to a constant at late times. It can be easily verified that this asymptotic constant is of order unity for the above range of observationally motivated $\langle t_* \rangle$ values. This late-time equality of $\langle t_* \rangle$ and $t_{\rm dyn}$ was a feature of the HS03 calculation: because they assumed $t_{\rm cool} \sim t_{\rm dyn}$,  haloes reach a steady state in which the supply of cold gas replenishes the gas that is converted into stars. This is the physical reason why extrapolating the HS03 model into the far future yields the non-physical result of eternal star formation. In contrast, we properly account for the evolution of the cooling time via equation~\eqref{eq:tcool_final}. The ever increasing cooling time asymptotically shuts down star formation, as we will show in \S\,\ref{sec:results}.

\subsection{Baryon mass fraction in haloes}
\label{sec:baryons}
 
In the previous sections, we derived expressions for the nSFR in a halo of a given virial temperature, obtaining results that depended on the baryon mass fraction in the halo. HS03 assumed that this was always equal to the cosmic baryon fraction $f_{\rm b}$, but we know that the baryon fraction in haloes is generally lower than $f_{\rm b}$ \citep[e.g.,][]{Crain_2007, Shull_2012}; in fact, this is one way of formulating the `missing baryon problem' \citep[see the review by][]{Bregman_2007}.

The circular velocity of galaxies and clusters in the local universe is observed to correlate with the baryonic mass of their parent haloes through the bTFR.
\cite{McGaugh_2010} considered a compilation of relevant observations for galaxies and clusters at $z\approx 0$ \citep{McGaugh_2005, Giodini_2009, Stark_2009, Trachternach_2009, Walker_2009}, and fitted the resulting bTFR with the power law
\begin{equation}
\label{eq:McGaugh_fit}
    \log_{10} M_{\rm b} = \alpha \log_{10} V_{\rm c} + \gamma \, ,
\end{equation}
where $V_c$ is the circular velocity, $M_{\rm b}$ is the baryonic mass, and $\alpha$ \& $\gamma$ are fitting parameters. The power law was broken in three intervals of circular velocity: $V_{\rm c}<20\, \rm km\, s^{-1}$, $20\, \rm km\, s^{-1}< V_{\rm c} < 350 \, \rm km \, s^{-1}$, and $V_{\rm c} > 350 \, \rm km \, s^{-1}$. Such ranges of circular velocity correspond to dwarf galaxies, spiral galaxies, and clusters, respectively. The picture that emerged from \cite{McGaugh_2010} was that galaxies in the first two intervals of $V_{\rm c}$ were baryon deficient (i.e., $M_{\rm b} < f_{\rm b} M$), but that the baryon mass fraction of clusters appeared to saturate at values consistent with the cosmic baryon fraction. Subsequent works performed similar analyses on larger samples of data, and updated the values of the best-fit parameters \citep{Zaritsky_2014, McGaugh_2015, Bradford_2016, Lelli_2016, Papastergis_2016, Uebler_2017}.

The bTFR was found to hold also at higher redshifts. \cite{Uebler_2017} considered two samples of galaxies and clusters, one at $z<0.9$ and the other covering $0.9<z<2.3$. In both samples, the bTFR could be fitted with a broken power law. \cite{Uebler_2017} found no evidence for a redshift evolution of the power-law slope $\gamma$ in haloes with circular velocity $V_{\rm c} <242 \, \rm km \, s^{-1}$. The slope was consistent with the value $3.75$ found by \cite{Lelli_2016}. On the other hand, the normalisation of the bTFR exhibited a variation up to a factor $\sim 1.6$, across the observed redshift range. 

In summary, there is a well documented correlation between the baryon mass and the circular velocity of haloes, which is related to their velocity dispersion, and hence to their virial temperature. There appears to be a critical  temperature, above which the baryon mass fraction takes the global value, but below which haloes are baryon deficient. While the slope of the bTFR seems to be independent of redshift, there are indications for a redshift evolution of the normalisation, implying an evolution of the critical $V_{\rm c}$ value. Recent numerical results also show modest evolution of the slope of the bTFR in the redshift range $0<z<1$, although the magnitude of the variations is different depending on precisely how the galaxy rotational velocity is estimated \citep{Glowacki_2020}.

We want to develop a simple analytic model for $f_{\rm b,\, halo}$ that captures these key features. We describe the main logical steps in this section, while presenting the details in appendix~\ref{sec:baryon_model}. The core idea of our model is that the distance from the centre of the halo up to which baryons are bound to the halo is set by the balance between the gravitational potential and the momentum transferred to the gas by supernova winds. The former is simply the Newtonian potential generated by the spherically symmetric power-law matter distribution as in equation~\eqref{eq:dens_prof}, while the latter is proportional to the stellar mass that follows from the gas cooling and gas consumption mechanisms described earlier in this section.

As discussed in appendix~\ref{sec:baryon_model}, there will be a critical distance $r_{\rm crit}$, beyond which winds unbind the gas. However, $r_{\rm crit}$ must be smaller than the cooling radius: in our model, stars only form within the volume swept by the cooling front. 
The condition $r_{\rm crit} < r_{\rm cool}$ translates into a condition on the virial temperature of the halo. Specifically, at any redshift $z$ there is a critical temperature $T(z)$ such that the gas content of haloes with $T<T_{\rm crit}$ extends only up to $r_{\rm crit}< r_{\rm cool}$. Within this critical radius, baryons are presumed to exist with the full global baryon fraction, $f_{\rm b}$. But the mean baryon fraction of the halo is lower, as it is the ratio between the baryonic mass within the critical radius and the virial mass of the halo. However, as the virial temperature increases beyond $T_{\rm crit}$, $r_{\rm crit}$ approaches the virial radius, and the halo baryon fraction tends to $f_{\rm b}$.

The redshift evolution of the critical temperature is determined numerically, and depends only on the internal properties of the halo and on the cooling function. Because the virial quantities are defined in terms of the critical density, $T_{\rm crit}(z)$ is effectively set by the evolution of the Hubble constant. Appendix~\ref{sec:baryon_model} shows that we match the slope of the bTFR observed by \cite{Lelli_2016} for $\eta=15/7 \approx 2.14$. For this value, we also match the critical temperature corresponding to the break of the bTFR in \cite{McGaugh_2010}. We stress that this is a non-trivial result, as there was in principle no guarantee of matching both the slope and break of the bTFR for the same value of $\eta$. Even more remarkably, $\eta=15/7$ implies a Kennicutt-Schmidt index of $1.4$, which is the value found by \cite{Kennicutt_1998} -- although he also highlights that any value between $\sim 1$ and $1.9$ would be reasonable, depending on the redshift and on the properties of the galaxies considered. Finally, as we show in appendix~\ref{sec:baryon_model}, our model is also in good agreement with the results of observations and simulations of the baryon mass fraction in the IGM both at $z\lesssim 2$ \citep{Wei_2019, Li_2020} and at higher redshift
\citep[see e.g. the review by][]{Meiksin_review}. It is satisfying that our simple modelling seems to capture all the physical features of the bTFR described earlier, and that it succeeds in reproducing very different data sets. 

\subsection{Fiducial parameters}
\label{sec:fiducial}

\begin{table*}
    \centering
    \begin{tabular}{cccc}
        \hline
        Quantity & HS03 & This work\\
        \hline
        $t_{\rm cool}$ & $t_{\rm dyn}$ & equation~\eqref{eq:tcool_final}\\
        $f_{\rm b\, halo}$ & $f_{\rm b}$ & equation~\eqref{eq:fbh_T} \\
        Density slope $\eta$ &  1.65 & 2.14 \\
        $\langle t_* \rangle$ $(\rm Gyr)$ & 1.4 & 2.39\\
        IMF $\beta$ & 0.1 & 0.21\\
         & from \cite{Salpeter_1955} with $(M_{\rm inf}, M_{\rm sup}) = (0.1, \, 40)\, \rm M_{\odot}$ & from \cite{Chabrier_2003} with $(M_{\rm inf}, M_{\rm sup}) = (0.1, \, 100)\, \rm M_{\odot}$\\
        \hline
    \end{tabular}
    \caption{Summary of key differences between HS03 model and our extended formalism.
    }
    \label{tab:differences}
\end{table*}

Having determined the functional form of $f_{\rm b, \, halo}(T, \, z)$, we are now fully equipped to solve equations \eqref{eq:nSFR} and \eqref{eq:M*_high-z} in order to obtain the nSFR at low and high redshift, respectively, and then compute the CSFRD via equation~\eqref{eq:CSFRD}. Before presenting our results, we summarise the key differences between our model and HS03 in Table~\ref{tab:differences}. The first crucial extension to HS03 formalism is that the cooling time is not assumed to be equal to the dynamical time, but is instead given by equation~\eqref{eq:tcool_final}. This enables us to account for the collapse time of haloes, and to predict the CSFRD into the far future of the Universe. The second important generalisation concerns the baryon mass fraction in haloes: whereas HS03 assume it to be equal to the cosmic baryon fraction for all haloes, in our case it is given by equation~\eqref{eq:fbh_T}, which accounts for the depletion of gas within haloes due to stellar winds and naturally reproduces the observed trend of the bTFR.

The other differences between HS03 and our model that are reported in Table~\ref{tab:differences} concern the choice of the fiducial values of the $\eta$ and $\langle t_* \rangle$ parameters. Our choice $\eta=2.14$ is motivated by matching the slope of the bTFR observed by \cite{Lelli_2016}. This value happens to simultaneously fit also the knee of the observed bTFR \citep{McGaugh_2010}, and the index in the Kennicutt-Schmidt relationship \citep{Kennicutt_1998}. Our fiducial value of $\langle t_* \rangle$ was chosen as the best-fit to \cite{Genzel_2010} observations of the Kennicutt-Schmidt relationship.

On the other hand, HS03 determined the optimal values of $\eta$ and $\langle t_* \rangle$ by fitting the predictions of their hydrodynamic simulations \citep{SH03_sims}. However, the latter was in turn tuned such that the simulations would reproduce the Kennicutt-Schmidt relationship. Therefore, our logic behind the choice of the fiducial $\langle t_* \rangle$ is similar to that in HS03.

We note that values in Table~\ref{tab:differences} do not represent the only physically sensible choice for the parameters of the model, though. Indeed, any value of $\langle t_* \rangle$ between $1.46 \, \rm Gyr$ and $3.87 \, \rm Gyr$ is consistent within $3\sigma$ with \cite{Genzel_2010} observations, and $2.07\lesssim \eta \lesssim 2.2$ ($1.92\lesssim \eta \lesssim 2.41$) match the slope of the bTFR measured by \cite{Lelli_2016} within $1\sigma$ ($3\sigma$). Furthermore, as we will show in \S~\ref{sec:parameters}, slopes of the gas density profiles in the range $1.65 \lesssim \eta \lesssim 2.5$ are all in reasonable agreement with observations of the CSFRD. Therefore, the ability of our model to match observations is not contingent on carefully chosen parameter values.

\section{Results and discussion}
\label{sec:results}

In \S\,\ref{sec:formalism} we presented the formalism of our model. In this section, we implement it and discuss the results. Unless otherwise indicated, hereafter we will set $\eta= 2.14$ and $\langle t_* \rangle = 2.39 \, \rm Gyr$, for the reasons discussed earlier. As for the cosmological model, we will consider a flat \LCDM universe with $\Omega_{\mathrm{m}}=0.315$, $\Omega_{\Lambda}=1-\Omega_{\mathrm{m}}=0.685$, $\Omega_{\mathrm{b}}=0.049$, $h=0.674$, $\sigma_8=0.81$, and $n_s=0.965$, with the usual definitions of the parameters. This cosmological model is consistent with the Planck-2018 results \citep{Planck18}.

In \S\,\ref{sec:nSFR} we will show our results for the nSFR within haloes of different fixed virial temperatures, which we will use to compute the CSFRD in \S\,\ref{sec:CSFRD_res}. In the remaining subsections, we will compare our results with observations, discuss the strengths and shortcomings of our model, and compare our results with those of HS03.

\subsection{Normalised star formation rate}
\label{sec:nSFR}

\begin{figure*}
	\includegraphics[width=\columnwidth]{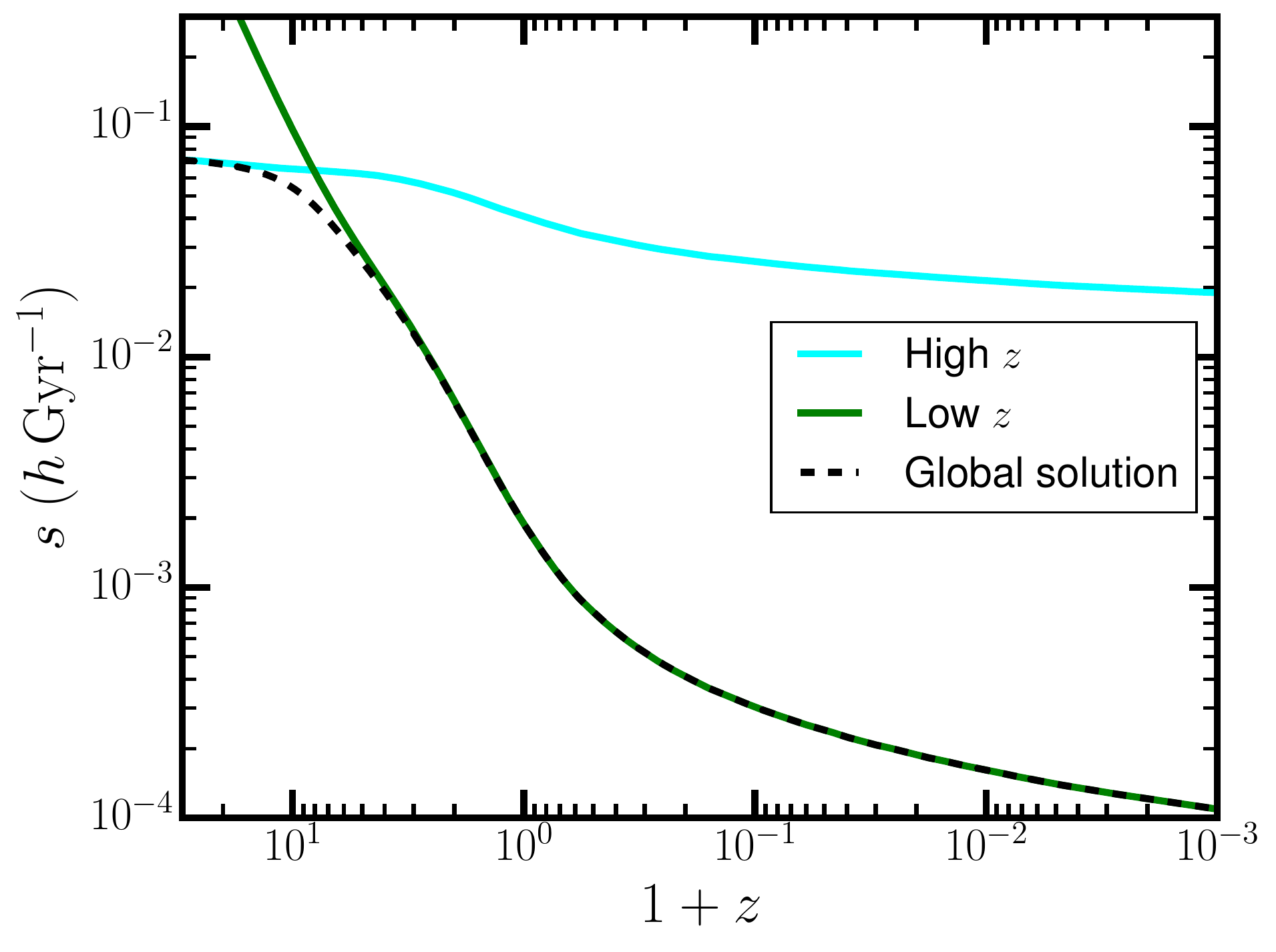}\hfill
	\includegraphics[width=\columnwidth]{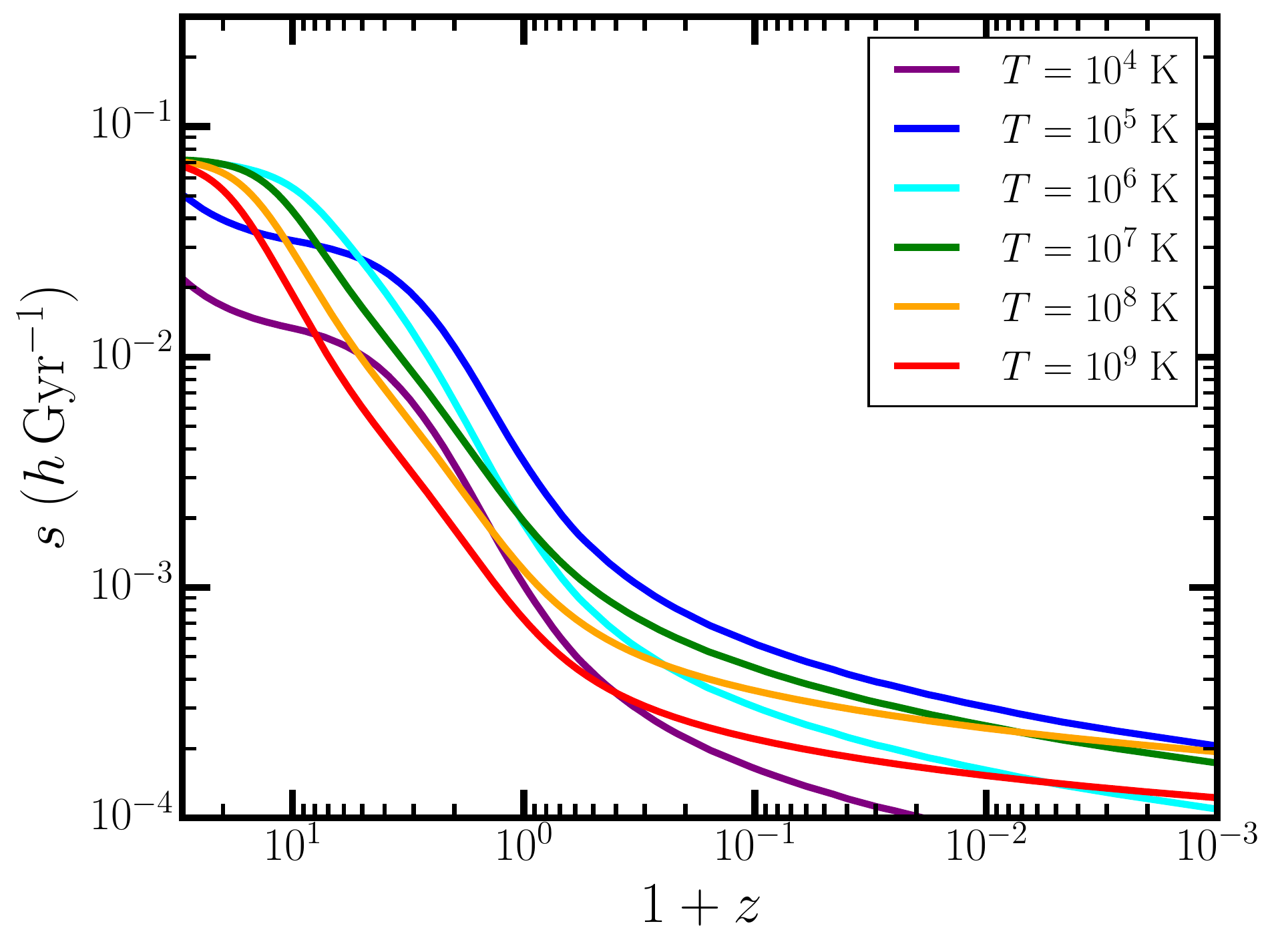}
    \caption{\textit{Left panel}: Numerical solution for the nSFR in the low-$z$ (solid green line) and high-$z$ (solid cyan line) regimes, via respectively equation~\eqref{eq:nSFR} or equation~\eqref{eq:M*_high-z}, for a halo with virial temperature $T=10^{6}\, \rm K$. The dashed black lines represents the nSFR over the entire redshift range obtained by connecting the high-$z$ and low-$z$ solutions in the aforementioned regimes with the smooth function in equation \eqref{eq:nSFR_smooth}. \textit{Right panel}: Normalised SFR for haloes with different virial temperatures. The solid purple, blue, cyan, green, orange and red lines correspond to $T=10^4\, \rm K$, $T=10^5 \, \rm K$, $T=10^6 \, \rm K$, $T=10^7 \, \rm K$, $T=10^8\, \rm K$ and $T=10^9\, \rm K$, respectively. In both panels, the slope of the gas density profile $\eta$ and the gas consumption time scale at high redshift $\langle t_* \rangle$ are fixed to the fiducial values reported in Table~\ref{tab:parameters}}
    \label{fig:nSFR}
\end{figure*}

We compute the nSFR in a halo of a given virial temperature $T$ via equation~\eqref{eq:nSFR} or equation~\eqref{eq:M*_high-z}, depending on whether we are in the low-$z$ or high-$z$ regime. As an example, we show the low-$z$ and high-$z$ solutions for a halo with virial temperature $T=10^{6} \, \rm K$ in the left panel of Figure~\ref{fig:nSFR} with solid green and cyan lines, respectively. We notice that the two lines intersect at a certain redshift, since the high-$z$ solution is nearly constant (not quite, owing to the changing baryon fraction), whereas the cooling-dominated low-$z$ solution evolves rapidly as a reflection of the changing mean density. We adopt a global solution for the nSFR by an empirical interpolation between these two limits:
\begin{equation}
\label{eq:nSFR_smooth}
    s(T,\,z) = \frac{s_{\rm high}(T,\,z) s_{\rm low}(T,\,z)}{(s_{\rm high}(T,\,z)^m+s_{\rm low}(T,\,z)^m)^{\frac{1}{m}}} \, ,
\end{equation}
where $s_{\rm low}(T,\,z)$ and $s_{\rm high}(T,\,z)$ are the low-$z$ and high-$z$ solutions given by equations~\eqref{eq:nSFR} and \eqref{eq:M*_high-z}, respectively. 

HS03 faced the same situation: at low $z$, the cooling-front argument from \S~\ref{sec:low-z} gave them an nSFR of the form \smash{$s(T, \, z)= \tilde{s}(T)q(z)$}, with \smash{$q(z) \propto \chi(z)^{9/2\eta}$}, where \smash{$\chi(z)=(H(z)/H_0)^{2/3}$}. But at higher redshift $q(z)$ had to exhibit an asymptotic behaviour, so  HS03 adopted the following interpolated expression:
\begin{equation}
\label{eq:q_z}
	q(z) = \left[ \frac{\tilde{\chi} \chi(z) }{ (\tilde{\chi}^m + \chi^m(z) )^{ \frac{1}{m} } } \right]^{\frac{9}{2\eta}} \, .
\end{equation}
Thus, in the above equation the HS03 parameter $\tilde{\chi}$ determines the redshift at which the transition between the high-$z$ and low-$z$ regimes occurs, while $m$ determines the smoothness of the transition.

There is a significant distinction between the two approaches to this transition. HS03 fine tuned $\tilde{\chi}$ to reproduce the nSFR resulting from simulations, whereas for us the transition is dictated by the value of $\langle t_* \rangle$, which is set by comparison to observations. The remaining nuisance parameter is the softening, $m$: we chose $m=2$, while HS03 set $m=6$. However, we verified that values of $m$ between $m=1$ and $m=10$ have negligible impact on the resulting CSFRD.

In the right panel of Figure~\ref{fig:nSFR} we show the global nSFR solution for haloes of different virial temperatures, between $10^4 \K$ and $10^9 \K$. At high redshift, all solutions with $T\geq 10^6\, \rm K$ converge to the same nSFR, as given by equation~\eqref{eq:M*_high-z}. The nSFR corresponding to $T=10^4 \, \rm K$ and $10^5 \, \rm K$ is however lower than in hotter haloes at high redshift. This happens because these temperatures are lower than the critical temperature above which the baryon mass fraction saturates to $f_{\rm b}$ (see \S\,\ref{sec:baryons} and appendix~\ref{sec:baryon_model}). The nSFR would increase up to the value found for the other virial temperatures at even higher redshift, but it is questionable whether star formation occurs at $z>30$ \citep{Abel_2002, Naoz_2006, Yoshida_2006, Gao_2007}.

\subsection{Cosmic star formation rate density}
\label{sec:CSFRD_res}

We now compute the CSFRD by integrating the nSFR over all virial temperatures, following equation~\eqref{eq:CSFRD}, where $s(T,\, z)$ is the nSFR computed as in equation~\eqref{eq:nSFR_smooth}. First we must consider the limits of the integration: following the argument made by HS03, we set the lower bound to $M_4(z)$, i.e. the virial mass corresponding to a halo with virial temperature $T=10^4 \, \rm K$. The logic behind this choice is that atomic line cooling is inefficient at lower temperatures if metals and molecular cooling are not considered. Whereas molecular cooling might be important at very early times, it is not expected to have a heavy impact on the global star formation history. We will discuss the impact of other possible choices for the minimum virial temperature of star-forming haloes in \S\,\ref{sec:parameters}. 

Regarding the upper bound of the integration, the exponential cutoff at the high mass end of the halo multiplicity function means that in practice the integration can be truncated at a suitably large finite virial temperature. We chose $T=10^{10}\,\rm K$ as an easily large enough maximum for all practical purposes.
To begin with, we therefore calculated the nSFR for haloes with temperatures spanning the range $10^4\, \mathrm{K} \leq T \leq 10^{10}\, \rm K$, in increments of $0.05 \, \rm dex$. Then, we obtained the function $s(T,\, z)$ through a numerical 2D-interpolation over the solutions in the temperature range considered.
We inserted the nSFR obtained in this way into the integral defining the CSFRD, performing the numerical integration with a simple trapezoid method, where we adopted an adequately fine integration step of $\Delta \ln (M/\rm M_{\odot})=0.1$.

We will show the results of the CSFRD for our fiducial choice of the parameters in \S\,\ref{sec:CSFRD_obs}. We will also show that another reasonable combination of the parameters underlying our model can yield an excellent agreement with observations. We will then discuss the detailed dependence of the CSFRD on the parameters in \S\,\ref{sec:parameters}.

\subsection{Comparison with observations}
\label{sec:CSFRD_obs}

\begin{figure}
	\includegraphics[width=\columnwidth]{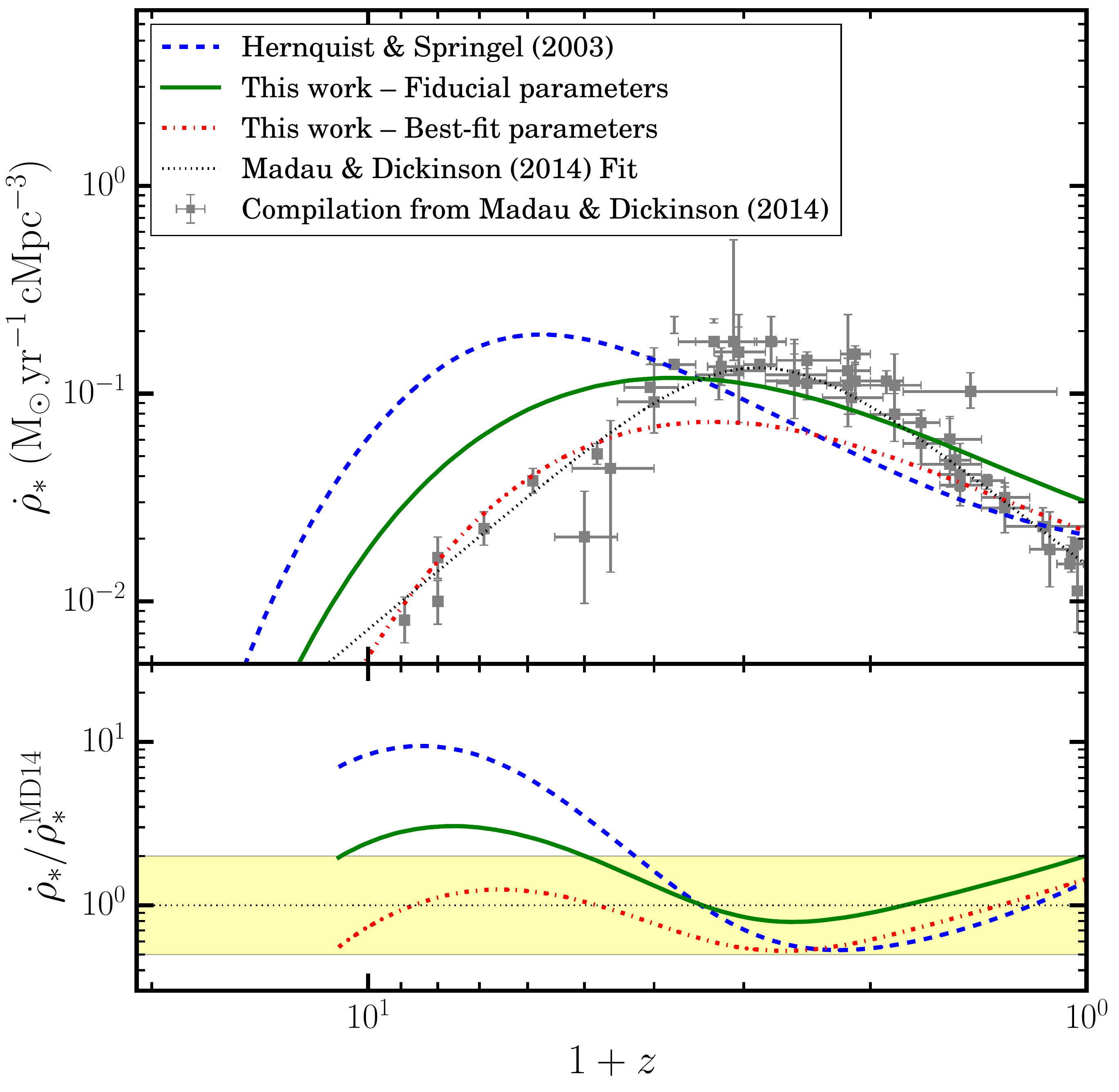}
    \caption{\textit{Top panel}: CSFRD predicted by our formalism with the fiducial and best-fit parameters in Table~\ref{tab:parameters} (solid green and dot-dashed red lines, respectively), and by the \protect\cite{HS03} model (dashed blue lines), compared with the compilation of observed data from \protect\cite{Madau_rev}. The black dotted line is the empirical fit to observations given by \protect\cite{Madau_rev}. \textit{Bottom panel}: Ratio between the CSFRD predicted by the models in the top panel (same colour coding and line styles) and the empirical fit to observations provided by \protect\cite{Madau_rev}. The horizontal black dotted line marks a ratio of $1$, to guide the eye. The shaded yellow region represents the deviations of the models within a factor of $2$ from the Madau--Dickinson fit. We do not plot the ratios for $z>10$ due to lack of observations. Our model is able to reproduce the CSFRD within a factor of $2$ in the entire redshift range considered.}
    \label{fig:CSFRD_obs}
\end{figure}

In the upper panel of Figure~\ref{fig:CSFRD_obs} we compare the results of our model with observations, and with the predictions of the original HS03 formalism. The data points represent the compilation of observations of the CSFRD provided in the review article by \cite{Madau_rev}. The dotted black line corresponds to the empirical best-fit curve to the data proposed by \cite{Madau_rev}. The dashed blue line refers to the CSFRD predicted by HS03, with the same choice of parameters as in their work. This is summarised in Table~\ref{tab:parameters} (fourth column). The green line represents the prediction given by our extended formalism, with our fiducial parameters (second column in Table~\ref{tab:parameters}). We also varied all parameters within a physically reasonable range (see also \S\,\ref{sec:parameters}), and identified a set of values that yield an excellent match with observations within the error bars (third column in Table~\ref{tab:parameters}). The corresponding prediction is the red dot-dashed curve. We caution that although we labelled this model `best-fit parameters', we did not carry out an exhaustive exploration of the parameter space: the purpose of the upper panel is simply to show that a range of parameters can provide a highly satisfactory match with observed data.

\begin{table}
    \centering
    \begin{tabular}{cccc}
        \hline
        parameter & fiducial & best-fit & HS03\\
        \hline
        $\eta$ &  2.14 & 1.9 & 1.65\\
        $\langle t_* \rangle$ $(\rm Gyr)$ & 2.39 & 3.87 & 1.4\\
        $\beta$ & 0.21 & 0.21 & 0.1\\
        $f_{\rm dyn}$  & 1 & 1 & --\\
        $T_{\rm min}$ & $10^4 \, \rm K$ & $10^{4.5} \, \rm K$ & $10^4\, \rm K$\\
        $m$ & 2 & 2 & 6\\
        $\tilde{\chi}$ & -- & -- & 4.6\\
        \hline
    \end{tabular}
    \caption{\textit{First column}: parameters used in our formalism or in the HS03 model. \textit{Second column}: Fiducial values of the parameters adopted in our model. \textit{Third column}: values of the  parameters of our model that best match the compilation of observed data provided by \protect\cite{Madau_rev}. \textit{Fourth column}: values of the parameters adopted by HS03. A value
    of `--' indicates a parameter that is not defined
    for a particular model.
    }
    \label{tab:parameters}
\end{table}

The accuracy with which observations are reproduced by our alternative models and by HS03 is quantified in the lower panel of Figure~\ref{fig:CSFRD_obs}, where we show the ratio between the predicted CSFRD and the Madau--Dickinson fit. For the fiducial parameters, we can reproduce the Madau--Dickinson fit within a factor of $2$ (shaded yellow region) for $z\leq 4$. This is rather satisfying, given the simplicity of the formalism. At higher redshifts, the discrepancy increases to a factor of 3 around $z= 6$. If we switch to the best-fit parameters, we can match the observations within a factor of 2 in the entire $0<z<10$ redshift range. However, it is not clear how concerned we should really be about discrepancies with the high-redshift CSFRD data, which are arguably less certain due to the inherent difficulties with UV observations.

It is noteworthy that we predict the peak of star formation at redshift $z\approx 2-3$, in good accord with the observations. Conversely, the peak appears at $z\approx 5-6$ in the HS03 model, in strong tension with observations. There are two key differences with HS03 that allow us to obtain a much better prediction for the peak of star formation. First, our choice of the gas consumption time scale is informed by observations of the Kennicutt-Schmidt relationship and do not rely on the results of specific cosmological simulations. 
Second, our model for the baryon mass fraction in haloes reduces the SFR in low-mass haloes at early redshift, which are also the haloes that dominate the halo mass function. In contrast, HS03 assumed that all haloes contained the cosmic fraction of baryons.

But it is not surprising that at low redshift our model gives similar result to HS03. Within our formalism the cooling time is of the order of the dynamical time in this regime (see Figure~\ref{fig:tcool}), so that the nSFR computed via equation~\eqref{eq:nSFR} does not considerably deviate from the nSFR calculated through the original HS03 formalism. However, the two calculations part company in the longer term, as we now discuss.

\subsection{Asymptotic behaviour}

\begin{figure}
	\includegraphics[width=\columnwidth]{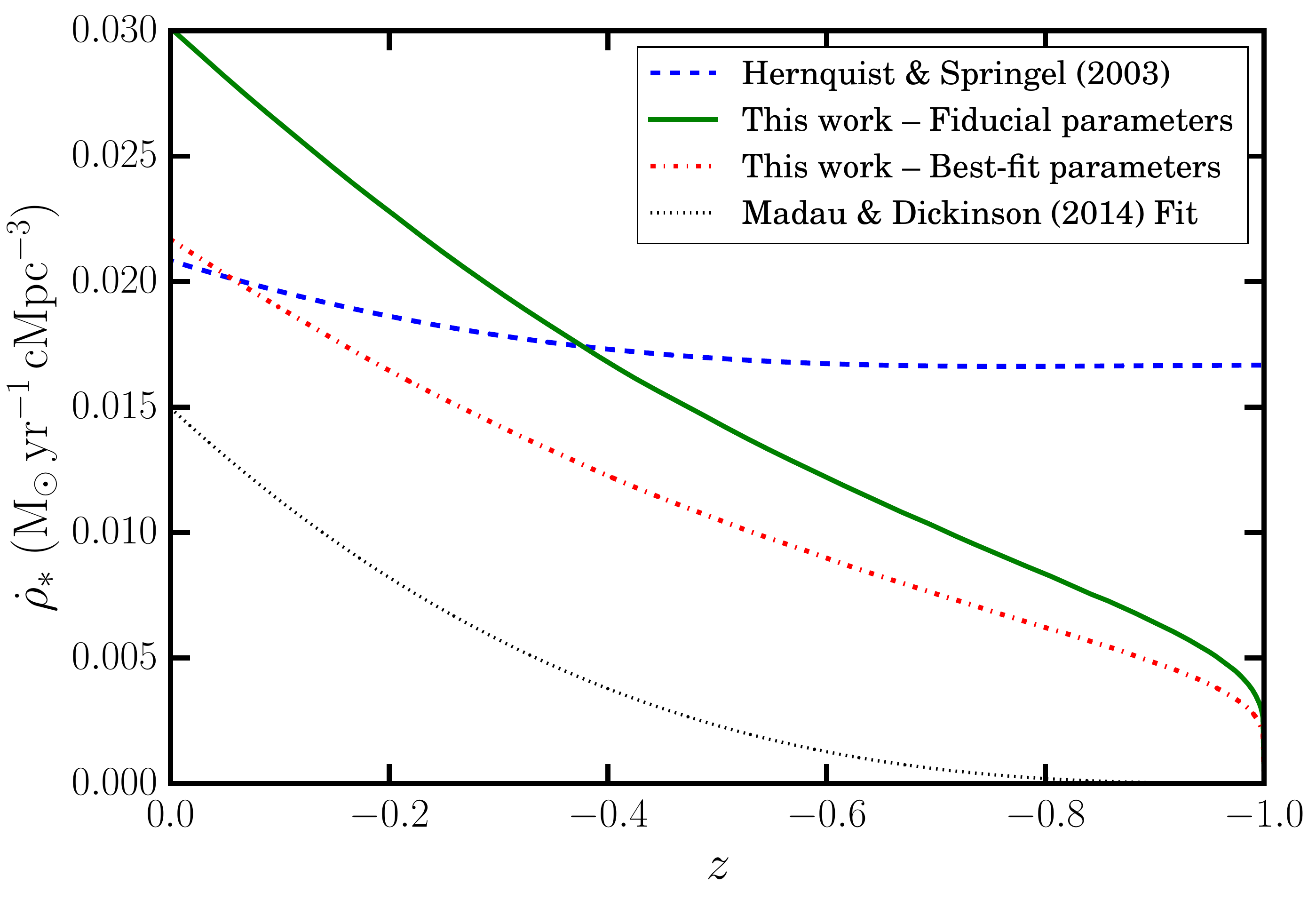}
    \caption{Future CSFRD (i.e., for $z<0$) predicted by HS03 formalism and our models with the fiducial and best-fit parameters in Table~\ref{tab:parameters}. We use the same colour coding and line styles as in Figure~\ref{fig:CSFRD_obs}. The black dotted line represents the extrapolation to $z\rightarrow -1$ of the fit by \protect\cite{Madau_rev} to observations of CSFRD. While the HS03 model converges to a non-null constant for $z\rightarrow -1$ (hence, $t\rightarrow \infty$), our model converges to zero. Therefore, unlike HS03, our model predicts a convergent star formation density over the full history of the Universe (see main text for further details).
    }
    \label{fig:CSFRD_theory}
\end{figure}

As we have seen, the HS03 assumption of $t_{\rm cool}=t_{\rm dyn}$ is physically problematic, because it predicts unending star formation, and thus a divergent total cosmic stellar density. Our model should give a more realistic prediction, as it allows $t_{\rm cool}$ to become arbitrarily large in the distant $\Lambda$-dominated future: thus the nSFR, and hence the CSFRD, should tend to zero. If we integrate the CSFRD over the full history of the universe (i.e., from $t=0$ to $t\rightarrow \infty$), we can therefore hope to obtain a finite cosmic star formation density; this represents a major improvement of our model over HS03. The radically different long-term behaviour of the two approaches can be seen in Figure~\ref{fig:CSFRD_theory}, which adopts the same colour coding as in Figure~\ref{fig:CSFRD_obs} but uses a scale linear in redshift for the $x$-axis so that we can show the full future history of the universe up to $z\rightarrow-1$. Strikingly, while the CSFRD obtained using the original HS03 method asymptotes to a constant shortly after the present time, the CSFRD given by our model continues to decay. 

Conversely, an extrapolation of the Madau-Dickinson fit into the future predicts a much steeper decline of the CSFRD. Obviously, there is no reason why such an unconstrained extrapolation should be considered reliable -- but it is interesting to note that this continues the trend seen at low positive redshifts, where the slope of the CSFRD predicted by our model is already somewhat shallower than the fit to the observations (see Figure~\ref{fig:CSFRD_obs}). Our predicted slope appears to be robust even to significant changes of the parameters of the model (Figure~\ref{fig:params}); thus at the current stage it is hard to identify the primary astrophysical reason behind the observed steepness of the CSFRD at low redshift. It is possible that by including AGN feedback mechanisms and (perhaps more importantly) by refining our modelling of the internal halo structure, gas cooling and stellar feedback, the CSFRD would decay more rapidly at low redshift (see the discussion in \S~\ref{sec:limitations}). We also point out that in the far future our predicted slope of the CSFRD becomes even shallower up to $z\approx -0.9$, after which star formation starts decaying faster again. This contrasts with a simple extrapolation of the Madau-Dickinson fit, which maintains the same steepness of the CSFRD from $z=0$ up to the infinite future. These differences have interesting implications for the asymptotic stellar mass produced in a unit volume throughout the history of the Universe.

\begin{figure*}
    \centering
    \includegraphics[width=0.49\textwidth]{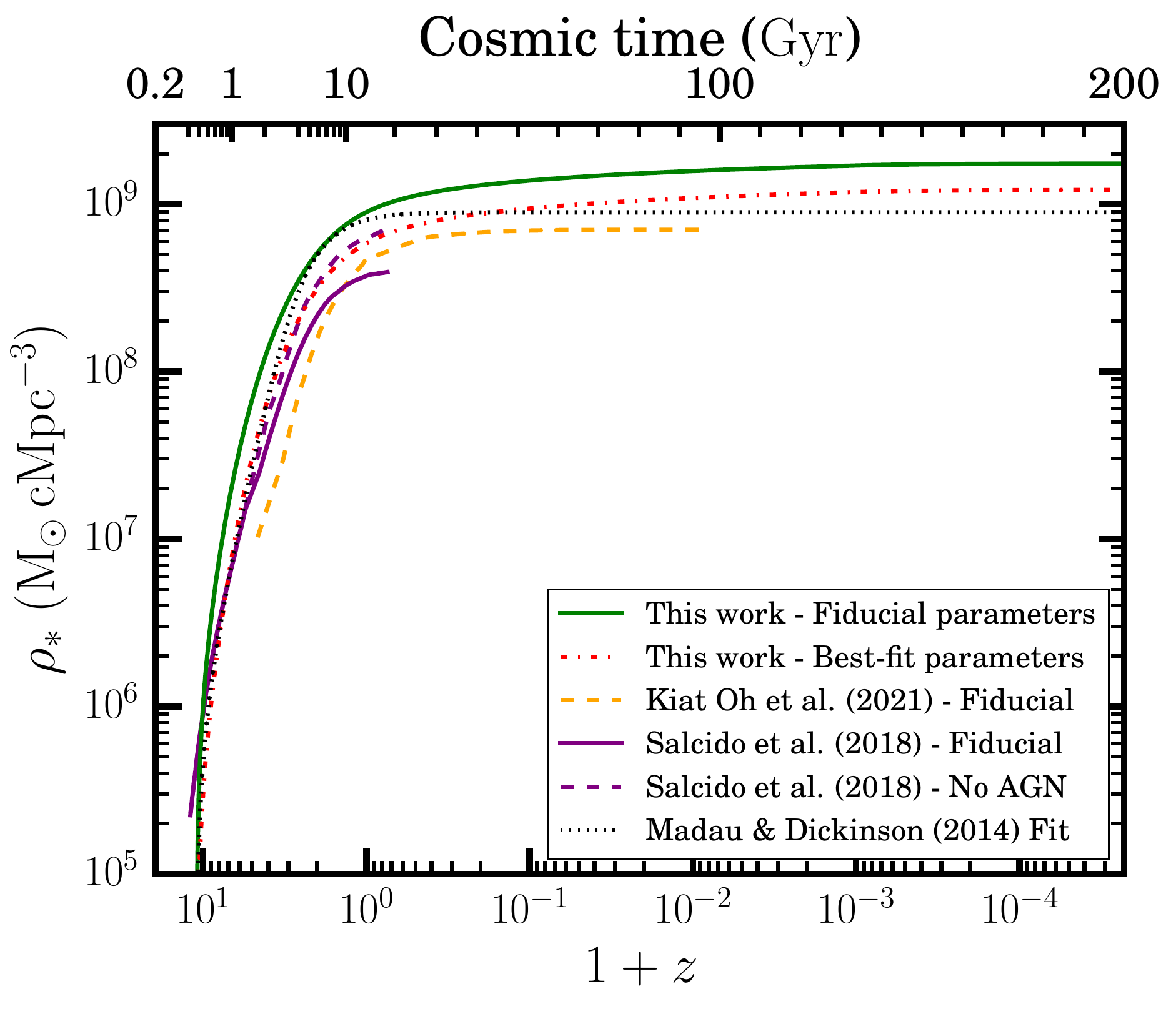}
    \includegraphics[width=0.49\textwidth]{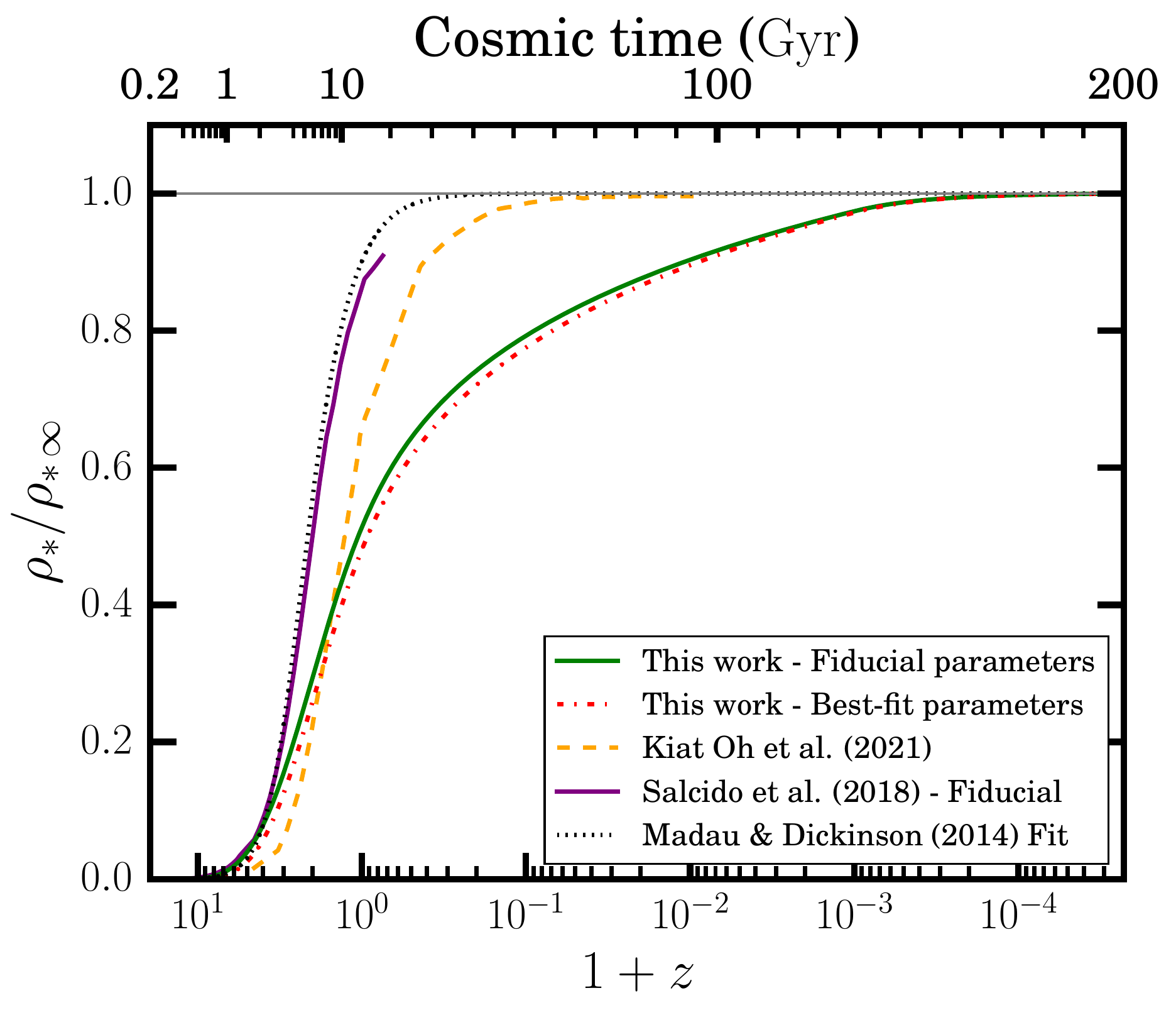}
    \caption{\textit{Left panel}: Cumulative stellar mass density produced up to redshift $z$. We show the results of our model for the fiducial and best-fit parameters, adopting the same colour coding as in Figures~\ref{fig:CSFRD_obs}-\ref{fig:CSFRD_theory}. We also plot the results of the simulation by \protect\cite{BK_2021} with the orange dashed line. The predictions of the fiducial EAGLE simulation and its no-AGN variant \protect\citep{Salcido_2018} are depicted with the purple solid and dashed lines, respectively. The black dotted line shows the stellar mass density fraction obtained by extrapolating the empirical fit by \protect\cite{Madau_rev} to observations of the CSFRD into the far future. \textit{Right panel}: Same as in the left panel, but normalised by the total stellar mass density formed throughout the history of the Universe. For the case of \protect\cite{Salcido_2018}, we follow their approach of estimating the asymptotic density via an  analytical extrapolation of the rapidly declining SFRD in their fiducial run from the end of their calculation at $20.7 \, {\rm Gyr}$ until $100\,{\rm Gyr}$ after the Big Bang; but no such convergence is apparent in their no-AGN run.}
\label{fig:SMD}
\end{figure*}

We can integrate the CSFRD predicted by our model from $z=10$, which is the highest redshift currently achieved by observations of star formation \citep{Bouwens_2012a, Bouwens_2012b}, up to a variable redshift $z$. This cumulative stellar mass density (SMD) at redshift $z$ is shown in the left panel of Figure~\ref{fig:SMD}. The asymptotic SMD for $z\rightarrow -1$ predicted by our model is  $\rho_{* \, \infty} = 1.7\times 10^9 \, \rm M_{\odot}\, cMpc^{-3}$ and $\rho_{* \, \infty} =1.2 \times 10^9 \, \rm M_{\odot} \, cMpc^{-3}$ for the fiducial and best-fit parameters reported in Table~\ref{tab:parameters}, respectively. If we instead consider the time-integral of the CSFRD over the redshift range where we do have observations, i.e. $0 < z< 10$, we obtain $9.0\times 10^{8} \, \rm M_{\odot}\, cMpc^{-3}$ and $5.8\times 10^{8} \, \rm M_{\odot}\, cMpc^{-3}$ in the fiducial and best-fit cases, respectively. By comparison, the time-integral of the Madau--Dickinson fit over the same redshift range is $8.0 \times 10^8 \, \rm M_{\odot}\, cMpc^{-3}$, and our fiducial (best-fit) model reproduces this figure within about 11\% (28\%). Thus, even though the best-fit parameters guarantee a better overall match with the Madau-Dickinson fit, the fiducial model actually yields a better agreement with the time integral of the CSFRD until $z=0$ (Figure~\ref{fig:SMD}, left panel).

We can also compute the fraction of the SMD formed up to a certain redshift with respect to the asymptotic value in the infinite future: this is shown in the right panel of Figure~\ref{fig:SMD}. With the fiducial and best-fit parameters, our model predicts that respectively $43\%$ and $39\%$ of the asymptotic total stellar mass density was already in place by $z=0.41$. This redshift is chosen to correspond to a look-back time of $\sim 4.5 \, \rm Gyr$, i.e. when the Sun formed. This is reassuring from the point of view of our typicality as observers. Similarly, the fraction of all stellar mass produced by the present time is $53\%$ and $49\%$ for the fiducial and best-fit parameters, respectively. By contrast, an extrapolation of the Madau-Dickinson fit all the way to $z\rightarrow -1$ would imply that $\approx 90\%$ of all stellar mass density that will ever be produced has already formed. This difference reflects the fact that our model predicts a slowly decaying CSFRD at negative redshift, as discussed above.

It is informative to compare the predictions of our model for the future star formation history with those of numerical simulations. For instance, \cite{Salcido_2018} ran a suite of cosmological simulations based on the EAGLE project \citep{EAGLE_Schaye2015} that extend about one Hubble time into the future. In the left panel of Figure~\ref{fig:SMD}, we report the evolution of the SMD that they obtained for a \LCDM cosmology consistent with Planck-2018 results \citep{Planck18} with their fiducial run and a variant without AGN feedback (solid and dashed purple lines, respectively). Our model is in excellent agreement with their no-AGN run up to $t\approx15\, \rm Gyr$, especially for the best-fit choice of the parameters, and we obtain a similar CSFRD at $z\sim 0$. On the other hand, our results are in tension with the far-future behaviour of the SMD in the fiducial run, which reaches a plateau much earlier than in our models. Thus the impact of AGN feedback in this model appears to truncate star formation more abruptly, which helps steepen the predicted CSFRD($z$) -- although in this case the resulting SMD is suppressed below the observed local value given by the Madau-Dickinson fit.
We also caution that \cite{Salcido_2018} could run their simulations `only' up to $20.7\, \rm Gyr$ since the Big Bang. The asymptotic stellar mass density for their fiducial model, $\rho_{*\infty}$, was estimated by extrapolating the data from the EAGLE run up to $10^4 \, \rm Gyr$ after the Big Bang. However, their no-AGN run is very far from convergence at the end of the calculation, so no meaningful estimate of $\rho_{*\infty}$ was possible in this case. Nevertheless, when we compare the \textit{fractional} SMD (right panel of Figure~\ref{fig:SMD}), we find clear tension between our model and the fiducial \cite{Salcido_2018} simulation. The amount of stellar mass density formed by $z=0$ in their fiducial run accounts for $\sim 88\%$ of the total stellar mass density over the full history of the Universe, which is almost twice as much as we predict.

The other interesting feature of the no-AGN run considered by \cite{Salcido_2018} is that the SMD does not reach a plateau. Rather, in this run the CSFRD decays very slowly after the present epoch and then exhibits an upturn and a revival of star formation up to a cosmic time of $20.7 \, \rm Gyr$ (see their figure 8; this feature is not so apparent in the integrated SFD shown in Figure~\ref{fig:SMD}). Such non-monotonic behaviour in the late-time SFRD, which is absent in our model, was interpreted by \cite{Salcido_2018} as a consequence of the lack of heating mechanisms in massive ($M_*>10^{11}\, \rm M_{\odot}$) galaxies, which would then no longer be able to resist gas cooling processes, hence allowing for new star formation.
A late-time revival of star formation was also predicted by \cite{BK_2021}, who computed the CSFRD up to $97 \, \rm Gyr$ after the Big Bang, with a suite of simulations based on the adaptive-mesh-refinement code \texttt{Enzo} \citep{Enzo_Bryan2014}. The various simulations included stellar feedback prescriptions based on \cite{BK_2020}, but no AGN feedback, and differed by mass resolution and level of mesh refinement. The authors found that the upturn in the CSFRD in the future occurred at cosmic time $\gtrsim 50 \, \rm Gyr$, i.e. later than in the case of \cite{Salcido_2018}. However, the upturn was observed at later times for higher levels of mesh refinement, so this feature was believed to be a numerical artefact rather than a consequence of the removal of AGN feedback.

We include the SMD predicted by the fiducial simulation by \cite{BK_2021} in Figure~\ref{fig:SMD} (dashed orange line). We note that the bare value of the asymptotic SMD is lower than in our model, but agrees within a factor 2, and that the best-fit parameters yield a better match with \cite{BK_2021}. However, in \cite{BK_2021} the SMD reaches the plateau at $t \approx 50 \, \rm Gyr$, whereas in our models this occurs at $t\approx 150 \, \rm Gyr$. Nevertheless, our predictions of the fractional SMD (right panel of Figure~\ref{fig:SMD}) are in good agreement with \cite{BK_2021}, at least at $z>0$. In particular, at $z=0$ their fiducial simulation predicts that $\sim 60\%$ of the asymptotic stellar mass has already formed.  This is close to the figure obtained with our models ($49-53\%$, depending on the choice of the parameters), but quite in tension with the \cite{Salcido_2018} result (88\%). If we consider negative redshifts, then our model departs from \cite{BK_2021} too. This is probably due to a combination of our simplified picture of star formation, and of the finite resolution of the simulations. We plan to undertake a rigorous comparison of our assumptions against cosmological simulations in future work. Regarding numerical resolution, \cite{BK_2021} argue that it could have an important impact especially on the late-time SMD, with some of their runs predicting a fractional SMD at the present time as low as 40\%.

To summarise, the asymptotic behaviour of our model is in good agreement with the no-AGN EAGLE run \citep{Salcido_2018} up to about 15\,Gyr, although the results start to diverge at the end of their calculation. Our work provides an SMD history that is in broad agreement with the numerical results by \cite{BK_2021}, especially regarding the fractional SMD up to present time. In the far future, all works considered in this section are in tension with one another. From a physical point of view, the discrepancies of our predictions with numerical work (and with the slope of the Madau-Dickinson curve for $0<z<2$) may be alleviated by addressing the simplifications in of our modelling (see \S~\ref{sec:limitations}). From a numerical standpoint, running high resolution simulations extended out to several Hubble times into the future seems crucial to achieve reliable predictions. Although feasible, this demands heavy computational resources under current computational constraints. Thus, a fuller understanding of the future of star formation may well require a suitable combination of analytic and numerical models. With this point in mind, we note that the fractional SMD given by our model appears to be robust even to large changes of model parameters (see Table~\ref{tab:parameters}). This makes our approach particularly suitable for revealing any impact of cosmological parameters on the star formation history, and in future work we therefore plan to use our code as a test-bed for anthropic arguments in resolving cosmic coincidences.

\subsection{Dependence on the parameters of the model}
\label{sec:parameters}

\begin{figure*}
    \centering
    \includegraphics[width=\textwidth]{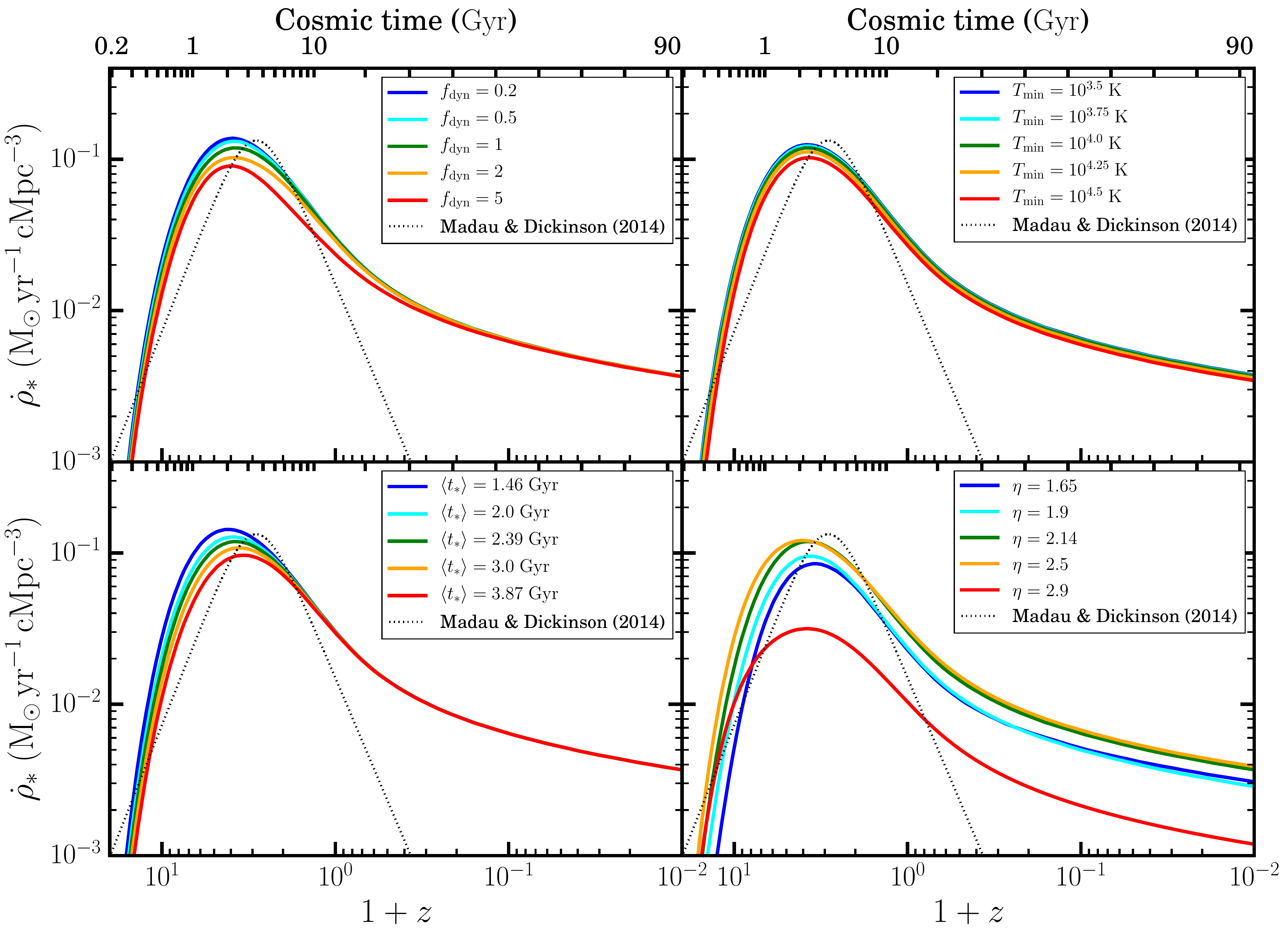}
    \caption{
   We investigate the effect of the various parameters of our model on the predicted CSFRD. Proceeding clockwise from the upper left panel, we show the impact of: the multiple of the dynamical time to which the cooling time converges at early times (see equation~\eqref{eq:tcool_smooth} and the relative discussion in \S\,\ref{sec:low-z}); the minimum virial temperature of star-forming haloes; the slope of the gas density profile $\eta$; the gas processing time scale $\langle t_* \rangle$. In every panel, the lower axis shows $1+z$ and the upper axis the corresponding cosmic time for the cosmological parameters considered in this work (consistent with Planck-2018 results, see \S~\ref{sec:results}). The solid lines represent the predicted CSFRD; different colours correspond to different values of the parameter being studied, as reported in the respective legend. All other parameters are fixed to the fiducial values in Table~\ref{tab:parameters}. As a comparison, we also report the empirical fit to observed data provided by \protect\cite{Madau_rev}. 
    }
    \label{fig:params}
\end{figure*}

It is interesting to look in more detail at the robustness of our model predictions, and
Figure~\ref{fig:params} examines the dependence of the CSFRD on the parameters of our model. In each panel, we show the CSFRD resulting from varying only one of the parameters, while leaving all the others fixed to the fiducial values reported in Table~\ref{tab:parameters}. 

In the upper left panel of Figure~\ref{fig:params} we investigate the impact of $f_{\rm dyn}$ on the CSFRD; the lines are colour coded according to the value of $f_{\rm dyn}$. As expected, this parameter has a visible impact only at low and positive redshift, while in the far future of the Universe the CSFRD converges to the same solution, regardless of $f_{\rm dyn}$. This is a direct consequence of our definition of the cooling time given in equation~\eqref{eq:tcool_smooth}: $t_{\rm cool}$ tends to $f_{\rm dyn} t_{\rm dyn}$ at early times, whereas at late times it is roughly equal to the age of the Universe. Even in the redshift range where the effect of $f_{\rm dyn}$ is appreciable, its impact is largely sub-dominant with respect to that of the gas consumption time scale or of $\eta$. This is a reassuring result, considering that there is no strong motivation to pick one particular value of $f_{\rm dyn}$, which can be considered a nuisance parameter. The only case in which the impact of $f_{\rm dyn}$ seems to be comparable with that of $\langle t_* \rangle$ is $f_{\rm dyn} =5$, indicating that this may already be an unreasonably large value.

In the upper right panel, we study how the lower bound of the integral in equation~\eqref{eq:CSFRD} affects the final results. Indeed, as discussed in \S\,\ref{sec:CSFRD}, setting $M_4(z)$ as the minimum mass of star forming haloes has some physical justification, but in principle other choices could be made. The different lines span a range of solutions where the minimum virial temperature of star-forming haloes was varied between $T_{\rm min}=10^{3.5}$ and $T_{\rm min}=10^{4.5}\,\rm K$, respectively. Obviously, a lower $T_{\rm min}$ results in a higher normalisation of the CSFRD. We notice that while the values of $T_{\rm min}$ that we consider here span one order of magnitude, the differences in the resulting CSFRDs are at most within a factor of $\approx 1.5$. As we shall now see, these changes are sub-dominant with respect to the effect of the other astrophysical parameters of the models.

We first study the effect of the gas consumption time scale. This parameter impacts the redshift of the peak of star formation in a straightforward manner. Indeed, the peak is a consequence of the transition between the high-$z$ and low-$z$ regimes. Such a transition is determined by the gas consumption time scale $\langle t_* \rangle$, as it regulates the asymptotic nSFR at high redshift (see \S\,\ref{sec:high-z}). Thus, we would expect the peak of the CSFRD to occur at lower redshift for larger values of $\langle t_* \rangle$, in concordance with equation~\eqref{eq:M*_high-z}. This is exactly the behaviour that we observe in the lower left panel of Figure~\ref{fig:params}, where every line corresponds to a different value of the gas consumption time scale. It is apparent from this plot that the peak of the CSFRD appeared at excessively high redshift in HS03 largely because of their choice of the average gas consumption time scale ($\langle t_* \rangle = 1.4 \, \rm Gyr$, very close to the value corresponding to the blue line in Figure~\ref{fig:params}).

The value of $\eta$ has the most significant impact on the final CSFRD, in a manner that appears more complex than in the previous cases (lower right panel of Figure~\ref{fig:params}). This is not surprising, because both the nSFR at low redshift and the baryon mass fraction at all redshifts depend on $\eta$ in a non-trivial way, as can be seen in equations~\eqref{eq:nSFR}-\eqref{eq:SFR_final}, and by observing the trend of the critical temperature in Figure~\ref{fig:Tcrit}, which is obtained by solving equation~\eqref{eq:Tcrit_cond}. In the far future, varying $\eta$ mostly affects the normalisation of the CSFRD, while leaving the slope almost unchanged. The reason is that as $z\rightarrow -1$ the critical temperature exhibits nearly the same trend within the range of $\eta$ considered, with $\eta$ effectively changing only the normalisation of $T_{\rm crit}(z)$ (see Figure~\ref{fig:Tcrit}).

At high redshift, the nSFR is mainly set by the baryon mass fraction in haloes, as dictated by equation~\eqref{eq:M*_high-z}. Because the baryon content within haloes below the critical temperature declines more steeply for lower values of $\eta$ (see Figure~\ref{fig:fbh_T} in the Appendix~\ref{sec:baryon_model}), the nSFR in such haloes is more strongly suppressed at high $z$. Thus, one would expect that at high redshift the CSFRD is larger for larger values of $\eta$. Whereas we do observe this trend for $1.65 \leq \eta \leq 2.5$, in the case $\eta=2.9$ the CSFRD can be almost one order of magnitude smaller than in the other cases.

\begin{figure*}
    \centering
    \includegraphics[width=\textwidth]{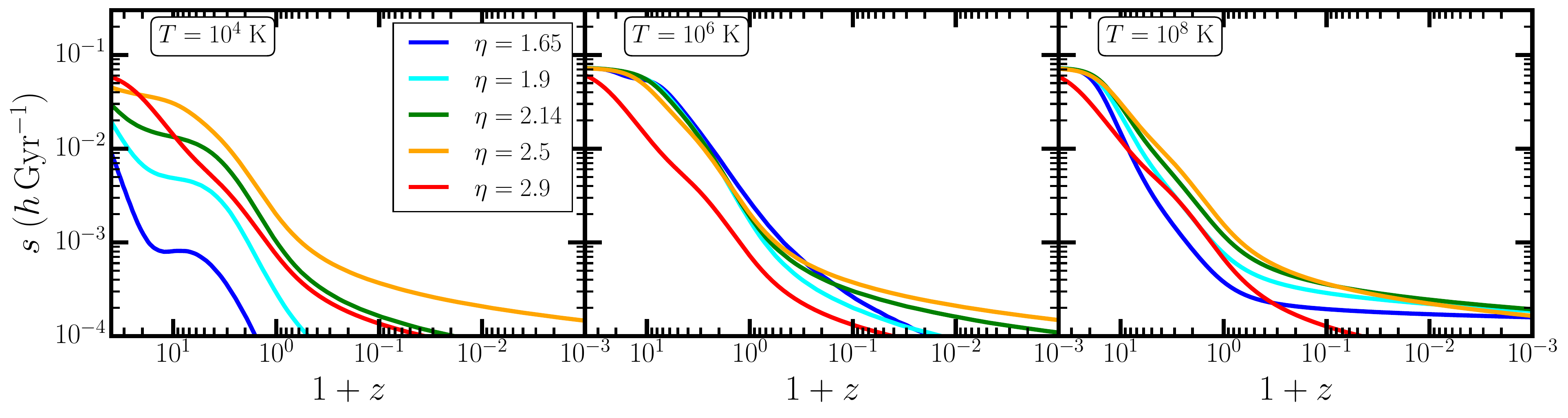}
    \caption{Impact of the slope of the gas density profile $\eta$ on the redshift-evolution of the normalised SFR within haloes of different virial temperatures. From left to right, each panel refers to haloes with $T=10^4 \, \rm K$, $T=10^6 \, \rm K$ and $T=10^8 \, \rm K$, respectively. In all panels, the lines are colour coded according to the value of $\eta$ considered, as reported in the legend within the left panel. The non-trivial trend of the normalised SFR at varying $\eta$ reflects the complex dependence of the halo baryon mass fraction and the low-$z$ prescription for star formation on this parameter (see equation~\eqref{eq:nSFR} and equations~\eqref{eq:fbh_T}-\eqref{eq:Tcrit_cond}).
    }
    \label{fig:nSFR_eta}
\end{figure*}

To gain a better understanding of this behaviour, it is best to start by examining the dependence of the nSFR on $\eta$. For this purpose, in Figure~\ref{fig:nSFR_eta} we show the redshift evolution of the nSFR  for a halo at a fixed virial temperature, for different slopes of the  gas density profile. From left to right, the three panels refer to haloes with virial temperature $T=10^4 \, \rm K$, $T=10^6 \, \rm K$ and $T=10^8 \, \rm K$, respectively. In all panels, the lines are colour coded according to the value of $\eta$ considered, as reported in the legend within the left panel.

The left panel confirms our expectations for the nSFR at high redshift: in low-$T$ haloes, star formation is more strongly suppressed for lower values of $\eta$. However, as we move to higher temperatures, this trend is no longer monotonic for a fixed high redshift, as is particularly evident from the $\eta=2.9$ case. The key point here is that the effect of $\eta$ on the overall nSFR is dictated both by the baryon mass fraction and the nSFR prescription at low redshift via equation~\eqref{eq:nSFR}. The latter actually becomes the dominant factor in shaping the nSFR for larger values of $\eta$. Indeed, as the slope of the gas density profile approaches the forbidden value $\eta = 3$, we have that $\tilde{S}(T) \rightarrow 0$ regardless of the virial temperature (see equation~\eqref{eq:SFR_final}), and consequently $s_{\rm low}$ is strongly suppressed. It follows that the asymptotic high-$z$ nSFR for larger $\eta$ can be much larger than the nSFR in the low-$z$ regime. Indeed, Figure~\ref{fig:nSFR_eta} shows that for $\eta=2.9$ the nSFR drops by $\sim 2$ orders of magnitudes from $z=20$ to $z=0$, and $\sim 3$ orders of magnitude at negative redshift. Thus, when we interpolate between the high-redshift and low-redshift solutions via equation~\eqref{eq:nSFR_smooth}, the resulting nSFR becomes very steep in the transition between the two regimes. As a result, the overall nSFR is suppressed in the entire redshift range considered, as visible in all panels of Figure~\ref{fig:nSFR_eta}. This feature is then of course reflected in the overall suppression of the CSFRD that we observed in the lower right panel of Figure~\ref{fig:params}. The peculiar behaviour discussed suggests that $\eta=2.9$ may not be a physically realistic choice. We included such value in our plots mostly to test our model at the boundaries of the permitted range of $\eta$. Indeed, $\eta=2.9$ is inconsistent with the observations considered to determine the fiducial parameters by more than $3\sigma$ (see \S~\ref{sec:fiducial}).

We conclude by noting that $\eta$ has a marginal effect on the location of the peak of star formation. Indeed, all values provide a redshift for cosmic noon in reasonable agreement with observational constraints ($2\leq z \leq 4$).

\subsection{Comparison with previous work}

In this subsection, we will discuss our results in the context of other previous related works. Recently, \cite{Salcido_2020} proposed an analytic model of star formation where the SFR within haloes is obtained from the efficiency with which baryons are converted into stars, which is assumed to depend only on the halo mass, and is parametrised as a broken double power law as in \cite{Moster_2018}. The indexes of the power laws below and above the critical break halo mass reflect the action of supernovae and AGN feedback. Besides the results of their fiducial model, these authors explore a number of variants. While we broadly agree with their fiducial model, \cite{Salcido_2020} seem to reproduce the peak of the CSFRD more closely, as well as the high-redshift observations. However, we remind the reader that our model neglects AGN feedback, and quite interestingly we agree at low redshift with the predictions of the no-AGN variant of the \cite{Salcido_2020} model.

\cite{Salcido_2020} also consider a model where the overall efficiency of star formation is $2.56$-times larger than their fiducial value, and obtain an earlier peak for the CSFRD that is compatible with our results for the fiducial parameters. This raises the question of whether our fiducial model might also display an excessive efficiency of star formation.
In part this concerns the average gas consumption time scale $\langle t_* \rangle$, which is a proxy for the star formation efficiency at high redshift. Our fiducial value of $\langle t_* \rangle = 2.39 \, \rm Gyr$ is set by matching the \cite{Genzel_2010} observations of the Kennicutt-Schmidt relationship -- although we showed in \S\,\ref{sec:CSFRD_obs}-\ref{sec:parameters} that $\langle t_* \rangle = 3.87 \, \rm Gyr$, which is still consistent with \cite{Genzel_2010} within $3\sigma$, would yield a much better match with the observed cosmic noon.  Adopting an even larger time scale (up to the factor 2.56 scaling considered by \citealt{Salcido_2020}), would certainly push the CSFRD to lower redshift, hence further improving the agreement with observations. But such a large change is firmly ruled out by observations.

In the end, our modelling approach is significantly distinct from that of \cite{Salcido_2020}, so that a direct confrontation of the SFR efficiencies of the two approaches is not so straightforward. The fundamental difference between the two approaches is that we do not assume any empirical parametrisation for the SFR, and we model it from first principles. But the fiducial parameters in the \cite{Salcido_2020} model are chosen to reproduce results from simulations or observations of quantities that are directly related to the CSFRD (such as the parametrisation of the mass-dependent star formation efficiency by \citealt{Moster_2010}), so a better match with data is to be expected. In our case, we choose our fiducial parameters only to reproduce observations other than stellar efficiency or the CSFRD. Thus, the CSFRD computed through our formalism is a genuine prediction, and it is remarkable that we can obtain an agreement with data within a factor of a few over a wide redshift range. This is also what distinguishes our approach from modelling in which empirical relations are fitted with analytic functions whose parameters are constrained to reproduce observations \citep[e.g.][]{Behroozi_2013_model, Lu_2014, Moster_2018, Behroozi_2019, Grylls_2019}.

But the fact that we can reproduce observations for a given set of physically motivated parameters is hardly a proof that our model contains all relevant physics. As we will further discuss in \S\,\ref{sec:limitations}, important physical processes such as metal cooling and AGN feedback are missing from our formalism. To exhaustively encompass the relevant physics, other approaches should be adopted, such as utilising semi-analytic models (SAMs) \cite[e.g.][]{Galform, Galacticus, Henriques_2015, Lacey_2016}. As we emphasised at the outset, however, the speed of such codes makes it challenging to explore the model parameter space thoroughly.

The next level of complexity is represented by cosmological hydrodynamic simulations. Obviously, simulations can provide much more information that goes well beyond global properties such as the CSFRD. On the other hand, their computational requirements are far more demanding than SAMs. In any case, it is noteworthy that most state-of-the-art simulations \citep[e.g.][]{EAGLE_Schaye2015, McCarty_2017, IllustrisTNG2018, Simba_Dave2019} can reproduce the observed CSFRD within a factor of a few, which is no more accurate than we achieved with our much simpler analytic approach.

\section{Limitations of our model}
\label{sec:limitations}

While our model is designed for rapid exploration of the cosmological parameter space, it is less suitable for investigating the dependence of the CSFRD on the astrophysical processes regulating star formation. Indeed, whereas our formalism does provide some insight on the qualitative effect of the average gas consumption time scale and of the slope of the gas density profile within haloes, the modelling of haloes itself is undoubtedly oversimplified. For instance, cosmological N-body simulations show that the matter density profiles within haloes follows an NFW profile \citep{NFW}, and not a pure power law, as is the case in HS03 and our model. Furthermore, we assume a spherically symmetric density profile in the entire region within the virial radius, without modelling galactic discs.

Our model includes two basic aspects of the physics of structure formation: the expansion of the universe and gas cooling. Mergers are implicitly taken into account via the halo mass function, even though there is no explicit reference to the redshift at which collapsed structures form. Likewise, although we do not explicitly model feedback processes from active galactic nuclei (AGN), we do include the effect of stellar winds in our model for the baryon mass fraction in haloes.

In terms of gaseous astrophysics, our model shares some limitations with HS03. We consider a cooling function relative to a H/He plasma with primordial abundances, so that only atomic/ionic cooling is included in the determination of the SFR at low redshift. But molecular cooling is believed to play an important role for star formation at high redshift ($z\sim 20-30$; see e.g. \citealt{Bromm_1999, Abel_2002}). In the context of star formation, this means that our results effectively neglect Population III stars \citep[e.g.][]{Carr_1984}. Also, a recent analytic model following the evolution of neutral hydrogen in damped \lya\ absorbers proposes that accreting atomic neutral hydrogen becomes molecular once it reaches the interstellar medium of the galaxy, and only a fraction of it forms stars, while the majority is ejected in the form of galactic outflows \citep{Theuns_2021}. This implies that ideally analytic models of star formation should distinguish between the atomic and molecular phases. Furthermore, our analytic model does not follow the co-evolution of star formation in galaxies and reionisation of the intergalactic medium. We hope to investigate these refinements to our formalism in future work.

The contribution of metal-line cooling to the cooling rate is also neglected. Following the reasoning in HS03, the inclusion of metal-line cooling should have only a mild impact on the normalisation of the CSFRD. If the mixing of metals with the IGM caused by galactic outflows is efficient, that should raise the CSFRD at low redshift, albeit only by about $20-30\%$ by $z=0$ \citep{HS03}. As HS03 point out in their work, metal-line cooling would shift the peak of star formation at slightly lower redshift (in their case, from $z\sim 5.5$ to $z\sim 5$), but that would not alter significantly the predicted curve of the CSFRD. We would thus expect that the inclusion of metal cooling in our model would produce a peak of star formation in even better agreement with data for the fiducial choice of the parameters.

The absence of AGN feedback mechanisms in our model is another aspect that merits further discussion. In particular, the question is whether the inclusion of AGN feedback could improve the match with the Madau-Dickinson fit at low redshift. Indeed, the presence of AGN-driven outflows has been shown to be crucial for effective prevention of hot-mode gas accretion into massive galaxies after cosmic noon, subsequently starving them of the gas necessary to fuel star formation \citep[e.g.][]{van_de_Voort_2011, Bower_2017}. As such, some form of AGN feedback has become a standard component of the modelling of galaxy formation, both in SAMs \citep[e.g.][]{Bower_2006, Lacey_2016} and numerical simulations \citep[e.g][]{EAGLE_Schaye2015, McCarty_2017, IllustrisTNG2018, Blank_2019, Simba_Dave2019}. 

The BAHAMAS simulation \citep{McCarty_2017} includes an AGN feedback mechanism based on the \cite{Booth_2009} model, in which a fraction of the rest energy of gas accreting onto black holes is transferred to the surrounding gas particles, thus increasing their temperature. The CSFRD predicted by the BAHAMAS simulation qualitatively resembles the observed trend, but the data are underpredicted at high redshift and overpredicted at low redshift. In particular, in the redshift range $0<z\lesssim 1$, the slope of the simulated CSFRD is very close to the one predicted by our model. However, such simulated results arise from a complex interplay of different physical processes, while our model includes only a few simple physical ingredients. Also, as pointed out by \cite{McCarty_2017}, the imperfect match with the observed CSFRD may derive from numerical issues, such as the calibration strategy  adopted for the parameters of the feedback model and the finite resolution of the simulation. In short, one cannot conclude that the inclusion of AGN feedback in our formalism would yield an unchanged late-time CSFRD, based only on \cite{McCarty_2017}.

Nevertheless, the results from the EAGLE simulations point in a similar direction. \cite{Salcido_2018} showed that the no-AGN EAGLE run exhibits a $\lesssim 0.5 \, \rm dex$ increase in the present-day CSFRD with respect to the fiducial run, and hardly any change in the redshift of the peak of the CSFRD or its slope at low redshift. This scenario is also supported by a recent analytic model of star formation \citep{Salcido_2020}, which suggests that removing AGN-driven effects would (mildly) affect the normalisation of the late-time CSFRD, but not its slope. By contrast, \cite{Salcido_2020} show that stellar feedback has a much stronger impact on the overall star formation history, with the removal of stellar feedback even erasing the peak of star formation. The authors argue that efficient stellar feedback in low-mass haloes is the main factor responsible for the appearance of the peak of star formation, which would be set by the gas consumption time scale at early times and the slowing growth rate of haloes at late times \citep[see also the discussion in][]{Salcido_2018}.

It is worth noting that both simulations discussed above adopt uniquely a thermal AGN feedback prescription, based on the same model \citep{Booth_2009}. Other cosmological simulations such as Illustris \citep{Illustris_V2014} include multiple AGN feedback modes. Fast-accreting black holes inject part of the radiative energy released by infalling particles in the surrounding gas (`quasar-mode'). Slowly-accreting black holes operate through `radio-mode' feedback, where AGN jets inflate hot and buoyant gas bubbles within in the halo, following the model by \cite{Sijacki_2007}. \cite{Vogelsberger_2013} showed that switching off both these AGN feedback mechanisms would increase the CSFRD at $z=0$ by about $0.7 \, \rm dex$, and that the slope of  the CSFRD after cosmic noon would be less steep. Furthermore, they showed that this suppression of the late-time CSFRD is primarily due to the radio-mode AGN feedback.

The Illustris AGN feedback model was refined in the successor simulation, IllustrisTNG \citep{IllustrisTNG2018}. In particular, the radio-mode feedback mechanism was replaced by a purely kinetic feedback model, in which AGN jets stochastically impart momentum to the surrounding gas. \cite{IllustrisTNG2018} showed that both the quasar-mode mechanism and the new  kinetic model drive a suppression of the stellar-to-total halo mass ratio in massive haloes ($M\gtrsim 10^{12} \, \rm M_{\odot}$) at $z=0$. Also, they  showed that removing AGN feedback altogether would produce a larger (by a factor of $\sim 4$) and somewhat less  steep CSFRD at low  redshift. \cite{Weinberger_2017} demonstrated that if one removes only kinetic AGN feedback from slowly-accreting black holes, then the slope of the CSFRD at $z<2$ becomes more moderate. This reflects the fact that after $z=2$ most black holes operate in the low-accretion regime (see the discussion in \citealt{Weinberger_2017}).

Simba \citep{Simba_Dave2019} is another cosmological simulation that incorporates jets among other AGN feedback prescriptions. In the case of Simba, AGN jets are fast, purely bipolar winds ejected along the direction of the angular momentum of the black hole. As in the case of Illustris and IllustrisTNG, jest are activated as the accretion rate of a black hole falls below a certain threshold. The Simba suite of simulations indicates that AGN-driven jets are primarily responsible for evacuating baryons from haloes with stellar mass $M_*\gtrsim 10^{12} \, \rm M_{\odot}$ at $z=0$ \citep{Appleby_2021}, while AGN activity is essentially negligible in this respect at $z>2$ (Sorini et al., in prep.).

To summarise, previous theoretical work suggests that the inclusion of some AGN feedback mechanism in our model may well yield a more rapid decline of the CSFRD  after $z=2$, giving an even better match with observations. In particular, it would appear that including AGN jets would be a promising  strategy. Using  several variants of the IllustrisTNG simulation, \cite{Terrazas_2020} found that galaxies where the accumulated black-hole-driven kinetic wind energy exceeds the binding energy of the gas within them exhibit a sharp decrease in their gaseous content and specific SFR. Because our model for the baryon content of haloes is already based on the balance between gravitational potential energy and the energy of outward stellar winds, these findings give some encouragement that our theoretical framework for stellar feedback could be extended to include AGN feedback, following similar energetic arguments.

Finally, we note that other modifications to our method discussed earlier may also help to improve the late-time behaviour of our  model. For example, distinguishing between atomic and molecular gas would give a more accurate cool gas budget, and hence a better estimate of the late-time SFR within haloes. Furthermore, when we account for the action of stellar winds diminishing the baryon mass contained in haloes, we do not consider their effect on the gas density or temperature profiles. Such changes would alter the cooling time, and hence the nSFR, at low redshift, thus affecting the late-time slope of the CSFRD. But at the current stage, we are satisfied with the approximate validity of our present analytic model of cosmic star formation. As we have seen, this reproduces current observations of the CSFRD with reasonable fidelity and makes interesting predictions for the future. The fact that it is able to do so despite its simplicity and small number of parameters should be considered a significant achievement of the approach.

\section{Conclusions and Perspectives}
\label{sec:conclusions}

In this work we developed an analytic model of cosmic star formation by extending the classic analysis by \cite{HS03}. The cosmic star formation rate density (CSFRD) at any fixed redshift is determined by modelling the star formation rate (SFR) in haloes, normalised by their mass, and then integrating over all possible halo masses, weighted by the halo mass function. In each halo, the SFR is set by the average gas consumption time scale at high redshift, and by the gas cooling time scale at low redshift. 

We extend the HS03 formalism in two main aspects. First, we give a physically motivated definition of the cooling time that is applicable at arbitrarily large cosmic times, effectively allowing us to predict the CSFRD in the far future ($t \rightarrow \infty$) of the Universe. Secondly, rather than assuming that all haloes contain the same fraction of baryons, we account for the dependence of the baryon mass fraction on the virial temperature of the halo and redshift. We do this by including a simplified model of stellar winds (following \citealt{Grudic_2019}), but without accounting for AGN feedback.

Our model depends only on two astrophysical parameters: the average gas consumption time scale at high redshift, and the slope of the gas density profile within haloes, which is assumed to follow a power law. These parameters are chosen so that our model reproduces independent measurements of the baryonic Tully-Fisher relation and of the Kennicutt-Schmidt relation. We stress that this is very different from adjusting our parameters in order to match observations of the CSFRD, which is instead what we want to predict. 

We computed the CSFRD within our formalism for several choices of the underlying parameters, and compared our predictions with observations, and with the results of HS03. Our main conclusions are as follows:
\begin{enumerate}
    \item With our fiducial values for the parameters of the model, we reproduce the fit to CSFRD observations by \cite{Madau_rev} within a factor of $2$ up to $z<4$, and a factor of $3$ up to $z<10$. With a different physically motivated choice of the parameters, we can reproduce the CSFRD within a factor of $2$ in the entire $0<z<10$ range. This level of agreement is comparable to that achieved by most cosmological hydrodynamic simulations.
    \item For physically reasonable values of the underlying parameters of our model, the peak of the CSFRD occurs in the range $2<z<4$, in good agreement with observations. Thus, our extended formalism improves on the HS03 prediction that the peak would occur at $z\sim5-6$.
    \item If we extrapolate the HS03 model towards $t\rightarrow\infty$, the CSFRD converges to an eternal constant, so that the time-integral of the CSFRD over the full history of the Universe would diverge. Within our formalism, the CSFRD decays to zero in the infinite future, and the integrated CSFRD is convergent. 
\end{enumerate}

It is remarkable that our model is able to reproduce the observed CSFRD despite its simplicity. However, there is still room for improvement. For instance, our model does not consider metal line cooling, and does not include feedback from AGN-driven winds or jets. Moreover, the density profile of haloes is approximated with a power law, while one would expect it to follow more complex profiles, such as an NFW profile. These generalisations to the formalism could be implemented in future work, but we feel that the model has significant value as it stands.

The convergence of the predicted CSFRD makes our formalism suitable for the investigation of observer-weighting selection effects on cosmic coincidences such as the small non-null value of the cosmological constant, and we expect to address this question in future work. But in any case, the impact of cosmological parameters on the star formation history is clearly an astrophysically interesting question that motivates the desire to calculate the behaviour in a wide range of counter-factual universes. We have developed a flexible, accurate and fast method to compute the CSFRD that allows exactly such an investigation to be carried out.

\section*{Acknowledgements}

We are grateful to the anonymous referee for constructive comments, which improved the quality of this manuscript. We thank Katarina Kraljic, Lucas Lombriser, Romeel Dav\'e, Tom Theuns and Avery Meiksin for helpful comments and discussions. The authors are supported by the European Research Council, under grant no. 670193.

\section*{Data availability}

No new data were generated or analysed in this article.\\

\noindent
The code that we developed to produce the results presented in this manuscript is undergoing further extension, and we plan to release it publicly as part of follow-up work. Until then, the code may be released upon reasonable request to the corresponding author.

%%%%%%%%%%%%%%%%%%%%%%%%%%%%%%%%%%%%%%%%%%%%%%%%%%

%%%%%%%%%%%%%%%%%%%% REFERENCES %%%%%%%%%%%%%%%%%%

% The best way to enter references is to use BibTeX:

\bibliographystyle{mnras}
\bibliography{extHS}

% Alternatively you could enter them by hand, like this:
% This method is tedious and prone to error if you have lots of references
%\begin{thebibliography}{99}
%\bibitem[\protect\citeauthoryear{Author}{2012}]{Author2012}
%Author A.~N., 2013, Journal of Improbable Astronomy, 1, 1
%\bibitem[\protect\citeauthoryear{Others}{2013}]{Others2013}
%Others S., 2012, Journal of Interesting Stuff, 17, 198
%\end{thebibliography}

%%%%%%%%%%%%%%%%%%%%%%%%%%%%%%%%%%%%%%%%%%%%%%%%%%

%%%%%%%%%%%%%%%%% APPENDICES %%%%%%%%%%%%%%%%%%%%%

\appendix

\section{Baryon mass fraction in haloes}
\label{sec:baryon_model}

In this appendix we present a simple analytic model for the baryon mass fraction within haloes, $f_{\rm b, \, halo}$, intended to improve on the assumption in HS03 that all haloes contain a baryon mass fraction equal to the global value, $f_{\rm b}$. Our approach bears some similarity to the work of \cite{Rasera_2006}, who developed an analytic model for the evolution of different baryonic phases within haloes, in turned inspired by the stellar wind prescription in \cite{SH03_sims}. The key feature in common between our model and that of \cite{Rasera_2006} is that baryons can be lost from a halo via the energy input from supernova-driven winds associated with star formation. Our focus is however on setting up the model in a general way that could apply to any cosmology and any epoch, whereas the work of \cite{Rasera_2006} is more specific to evolution in $\Lambda$CDM up to the present.

The central idea of our model is that the gas content of haloes is determined by a balance between the gravitational potential and the energy of the winds ejected by supernova explosions. If a gas parcel originally approached the halo from infinity, the minimum distance that it can reach depends on its total energy. Conversely, if the gas particle is already at the core of the halo, it may only escape if its energy is high enough such that it can overcome the gravitational potential well, and be driven outwards by the winds.

If we assume spherical symmetry, we can easily calculate an effective potential accounting for gravity and pressure forces. We can then investigate under which conditions a gas particle is bound to the halo. As we will show, for a certain range of virial temperatures there is a critical radius, smaller than the virial radius of the halo, beyond which pressure forces due to supernova winds overtake gravity. Thus, only gas particles within this critical distance are bound to the halo. This will allows us to estimate the baryon mass fraction retained by haloes of different virial temperatures.

Despite its simplicity, our model succeeds in explaining the dependence of $f_{\rm b, \, halo}$ on the virial temperature inferred from observations of the bTFR \citep{McGaugh_2010, Lelli_2016}. It also connects with the observed index of the Kennicutt-Schmidt relationship, and it predicts a value of the baryon fraction in the IGM at high redshift in reasonable agreement with observations.

\subsection{Effective potential}
\label{sec:potential}

Consider an isolated spherical halo of virial mass $M$, virial temperature $T$ and virial radius $R$ at redshift $z$. For convenience, we divide the matter within the halo into three categories: gas, stars and DM. In principle, all matter components may extend out to the virial radius $R$, and we will assume that they all follow the same profile up to a constant factor. 

It is common to neglect stresses from magnetic fields and cosmic rays, and so estimate the gas density profile by solving the equation for hydrostatic equilibrium (e.g. \citealt{BT2008}). If we impose a polytropic equation of state for the gas, then one possible solution for $\rho_{\rm gas}$ is given by \eqref{eq:dens_prof}. The corresponding total matter density profile is simply obtained from $\rho(r) = \rho_{\rm gas}(r)/ f_{\rm gas}$, implying $M_{\rm gas} = f_{\rm gas} M$.

Now consider the effect of supernova-driven winds. Following \cite{Grudic_2019}, we assume that the momentum injection rate per stellar mass formed into the gas within the halo is equal to the constant value $f_{\rm w} = 1000 \, \rm km \, s^{-1} / 40 \, \rm Myr$, irrespective of the properties of the ambient medium. This is supported by several models for standard stellar populations \citep{Leitherer_1999, Bruzual_2003, Agertz_2013, Hopkins_2011, Martizzi_2015, Kim_2015}. We also assume that winds are ejected in a spherically symmetric fashion. Therefore, the net force exerted by winds ejected by supernovae within a spherical shell with radius $r$ and thickness $\delta r $ on the surrounding gas is given by $f_{\rm w} (dM_{*,\, \rm young}/dr) \, \delta r$, where $M_{*, \rm young}$ is the mass of young stars, as those are also the more massive stars, which dominate feedback. Following the HS03 formalism, the wind force can thus be expressed as $f_{\rm w} \beta x f_{\rm gas} dM/dr \, \delta r$. Therefore, the acceleration of a gas parcel of thickness $\delta r$ and transverse area $\delta A$ at distance $r$ from the centre of the halo caused by stellar winds is:
\begin{equation}
    a_{\rm w} = \frac{1}{\rho_{\rm gas}(r) \delta A \delta r} f_{\rm w} \beta x f_{\rm gas} \frac{dM}{dr} \delta r \frac{\delta A}{4 \pi r^2} =  f_{\rm w} \beta x \, .
\end{equation} 
In other words, supernova winds transmit a constant radial acceleration directed outwards to gas parcels within the halo.

Because all forces involved (gravity, hydrostatic pressure force and winds-induced force) are central and spherically symmetric, we can associate a potential with all of them. The resulting effective potential within the halo is
\begin{gather}
\label{eq:potential}
\Phi (r) = 
    \begin{cases}
 \frac{\eta^2-4\eta+2}{\eta (\eta-2)_{\phantom{\bigstrut}}} \frac{GM}{R} \left(\frac{r}{R}\right)^{2-\eta} - \frac{\Lambda c^2}{6} r^2 - f_{\rm w} \beta x r + K   &(r< R)\\    
  - \frac{GM^{\phantom{\bigstrut}}}{r} - \frac{\Lambda c^2}{6} r^2 &(r>R),    
\end{cases}
\, ,
\end{gather}
where $K$ is a constant determined by imposing continuity at $r=R$. For $r\geq R$, the potential is given simply by the attractive Newtonian gravitational potential and the repulsive term due to the cosmological constant $\Lambda$ in the weak-field approximation of \LCDM \citep[e.g.][]{Nowakowski_2001}. The first term in the $r<R$ case is given by the combined action of gravity and the hydrostatic pressure force. The $\eta$-dependent prefactor stems from the relationship between the polytropic index of the equation of state of gas and the slope of the density profile. The quadratic term in $r$ for $r<R$ is again the repulsive potential of the cosmological constant, while the linear term in $r$ represents the contribution of supernova winds to the effective potential $\Phi$. 

It can easily be seen that the effects of the term containing the cosmological constant appear to be significant only on sufficiently large scales \citep{Nowakowski_2001}. For $\Omega_{\Lambda}=0.685$, the vacuum repulsion dominates over gravity for $r \gtrsim 5.3 (H(z)/H_0)^{2/3} R$; $r \gtrsim 4.6\,  R$ in the $\Lambda$-dominated future of the Universe. Thus we can safely ignore the repulsive term in the effective potential for the purpose of determining the baryon mass fraction inside a halo.

\begin{figure}
    \centering
    \includegraphics[width=0.49\textwidth]{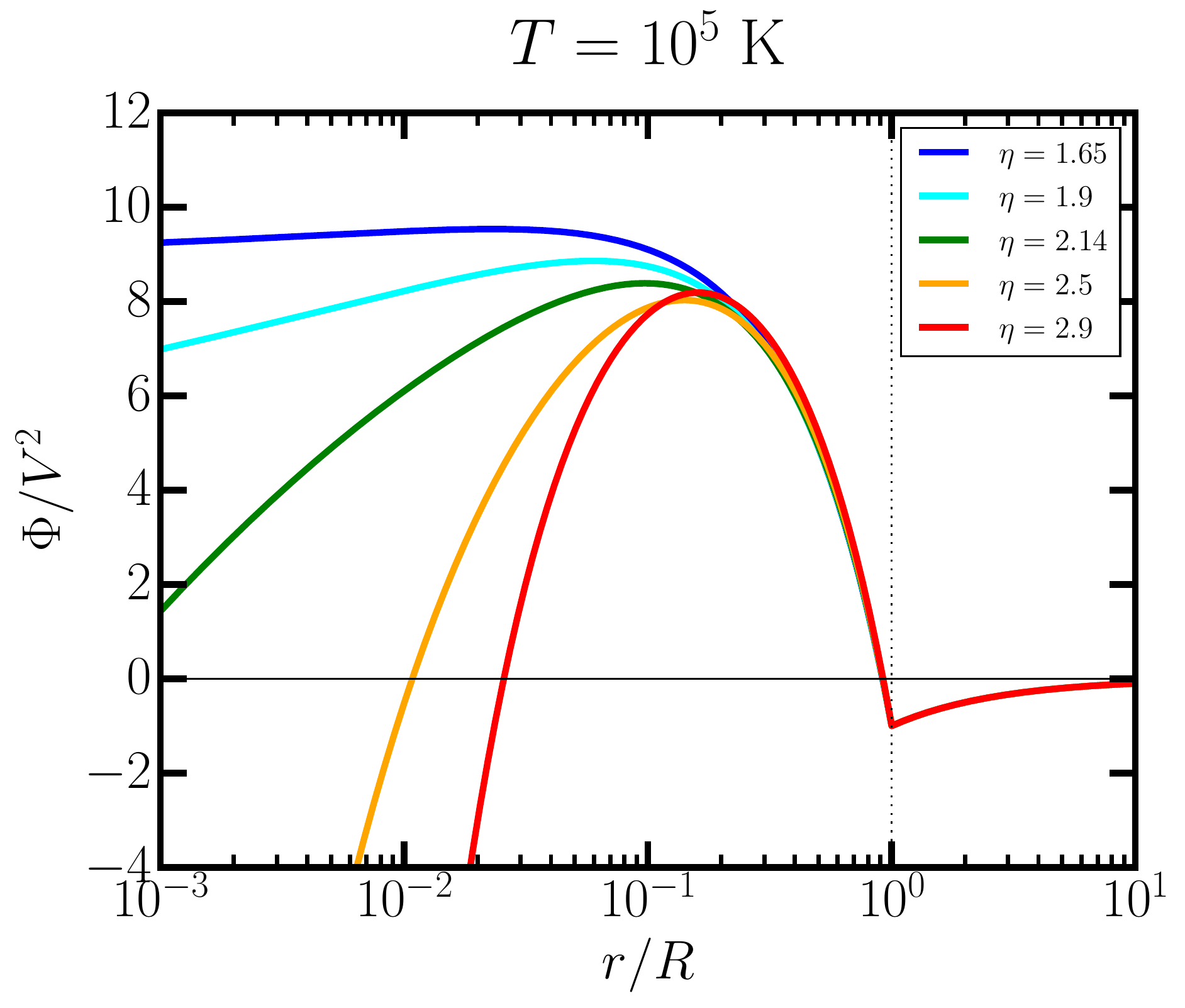}
    \caption{Effective potential $\Phi$, as given in equation~\eqref{eq:potential}, for a halo with virial temperature $T=10^5\K$ at $z=6$. The solid blue, cyan, green, orange, and red lines correspond to gas density profiles with slopes $\eta=1.65$,  $\eta=1.9$,  $\eta=2.14$,  $\eta=2.5$, and $\eta=2.9$, respectively. The vertical dotted black line and the horizontal thin black line mark the virial radius and $\Phi=0$, respectively, to guide the eye. The effective potential exhibits a maximum, meaning that only the gas within the distance corresponding to the maximum and with total energy lower than the potential barrier is bound to the halo (it may seem that the potential has a plateau for $\eta=1.65$, but there is in fact a maximum at $r/R\sim0.02$). As a result of the depletion of baryons beyond this peak in the potential, the mean baryon mass fraction within the halo is less than the cosmic baryon fraction.}
    \label{fig:potential}
\end{figure}

We point out that we implicitly assumed that the addition of stellar winds does not significantly affect the gas density profile given by equation~\eqref{eq:dens_prof}. However, equation~\eqref{eq:dens_prof} is a solution of the equation of hydrostatic equilibrium without any wind, so the expression for the effective potential is not fully self-consistent. But our aim here is to find the demarcation between low feedback that hardly alters the baryon content (in which case our solution is valid) and the point at which the baryon content is heavily reduced; our approach should still model this effectively.

Finally, we note that equation~\eqref{eq:potential} requires that $\eta \neq 0$ and $\eta \neq 2$. The case $\eta =0$ would correspond to a halo with constant density, which cannot give rise to any pressure gradient (assuming isothermality); thus for $r<R$ the potential will retain only the gravitational term and the linear term in $r$ due to stellar winds. This should be considered as a limiting case, rather than a realistic potential, because the relationship between the polytropic index of the equation of state of gas and the slope of the power-law gas density profile forbids $\eta=0$. Specifically, hydrostatic equilibrium with $T\propto \rho^n$ for a perfect gas admits the gas density profile given by equation~\eqref{eq:dens_prof}, with $\eta = 2/(1-n)$; $\eta = 0$ is therefore not possible. This condition guarantees that the cooling radius in equation~\eqref{eq:rcool} and the cooling rate in equation~\eqref{eq:cool_rate_gen} are well defined. On the other hand $\eta=2$ corresponds to the perfectly isothermal case. In this case, the gravitational potential is perfectly balanced by the hydrostatic pressure, and only the contribution due to stellar winds will be retained in equation~\eqref{eq:potential} for $r<R$.

It is instructive to visualise the shape of the effective potential for different values of $\eta$. In Figure~\ref{fig:potential} we now plot the effective potential as a function of radial distance, in units of the appropriate virial quantities, for haloes with virial temperature $10^5 \K$ at redshift $z=6$. We arbitrarily chose this relatively high redshift because the assumptions that led to equation~\eqref{eq:potential} are most exact in the high-redshift regime. 
Figure~\ref{fig:potential} shows that the potential is cuspy at $r=R$ for a wide range of values of $\eta$. Physically, this means that the acceleration of a test particle around $r=R$ is discontinuous. In other words, the cusp gives rise to a virial shock. The potential exhibits a maximum at a critical distance $r_{\rm crit}<R$ for all values of $\eta$ considered, so that there is an energy threshold for incoming gas to be able to reach the core of the halo. In particular, if a gas parcel has a total energy smaller than $\Phi (r_{\rm crit})$ \textit{and} it is located in the range $r_{\rm crit}<r<R$, it cannot overcome the potential barrier, and may be able to escape from the halo. Conversely, if the gas parcel in question has a total energy smaller
than $\Phi (r_{\rm crit})$ \textit{and} it is already within $r_{\rm crit}$, then it is bound to the halo. Therefore, the baryon mass fraction of the halo can be estimated as the ratio of the baryonic mass contained within $r_{\rm crit}$ and the total mass within the virial radius. 

However, the potential does not exhibit a maximum for all virial temperatures. At a given redshift one can identify a critical temperature above which the potential becomes monotonically increasing. In that case there is no potential barrier, and the mean halo baryon fraction is always equal to $f_{\rm b}$. In short, the physical behaviour depends in a complex way on the value of $\eta$ and on the virial temperature. We will give more detail on the different scenarios in the next section, focusing on those that are more relevant for this work.

\subsection{Temperature dependence of the baryon mass fraction}

As explained above, if the potential given by equation~\eqref{eq:potential} exhibits a local maximum at $r_{\rm crit} < R$, then the baryon mass fraction inside the halo $f_{\rm b,\, halo}$ will be reduced below the cosmic value $f_{\rm b}$. It is therefore important to understand for which values of $\eta$ and $T$ the condition $r_{\rm crit} < R$ is satisfied. But first, we need to recognise that there is an additional criterion to satisfy at low redshift. The main feature of the cooling-dominated regime of star formation explained in \S~\ref{sec:low-z} is that only the gas within $r_{\rm cool}$ will form new stars -- meaning that winds can be generated only within this radius. We can therefore use the potential in equation~\eqref{eq:potential}, but we will require that $r_{\rm crit} < r_{\rm cool}$ at low redshift. This means that in practice we need to understand under what conditions $r_{\rm crit} < \min(r_{\rm cool}, R)$ at any redshift. As we will show below, this criterion yields a good match with several observations related to the baryon mass fraction in haloes. Our approach thus treats the main physical features regulating the baryon content in haloes in a manner that respects the overall philosophy of the HS03 model, while also yielding realistic predictions.

If we study the behaviour of the effective potential given by equation~\eqref{eq:potential} for $r<\min(r_{\rm cool}, R)$, we can distinguish a number of different possible regimes:
\begin{itemize}
    \item $0< \eta \leq 2-\sqrt{2}$: The potential is monotonically decreasing regardless of the halo temperature, meaning that the gas tends to flow outwards, and the halo is unable to retain gas.
    \item $2-\sqrt{2} < \eta \leq 1.5$: For a sufficiently high virial temperature, the potential is monotonically decreasing, and hence the gas tends to escape from the halo. Below this temperature threshold, the potential exhibits a minimum at $r>0$, hence being able to retain gas. This is however a non-physical scenario, as it would predict that  massive haloes are more baryon-deficient than low-mass haloes, contrary to observations \citep[e.g.][]{McGaugh_2010}.
    \item $1.5< \eta <3$ and $\eta \neq 2$: The potential is monotonically increasing above a certain critical virial temperature threshold. Below the threshold, it exhibits a maximum at a critical distance $r_{\rm crit}$, given by
    \begin{equation}
        \label{eq:rb}
        r_{\rm crit} = \left( \frac{-\eta^2 + 4\eta -2}{\eta} \frac{GM}{f_{\rm w} \beta x R^2} \right)^{\frac{1}{\eta-1}} R \, .
    \end{equation}
    As discussed in the previous section, gas parcels within $r_{\rm crit}$ are bound to the halo in this case.
    \item $\eta =2$: The potential is linear in $r$ (see discussion in \S\,\ref{sec:potential}) and monotonically decreasing. Therefore, the halo cannot retain gas.
\end{itemize}

\begin{figure}
    \centering
    \includegraphics[width=\columnwidth]{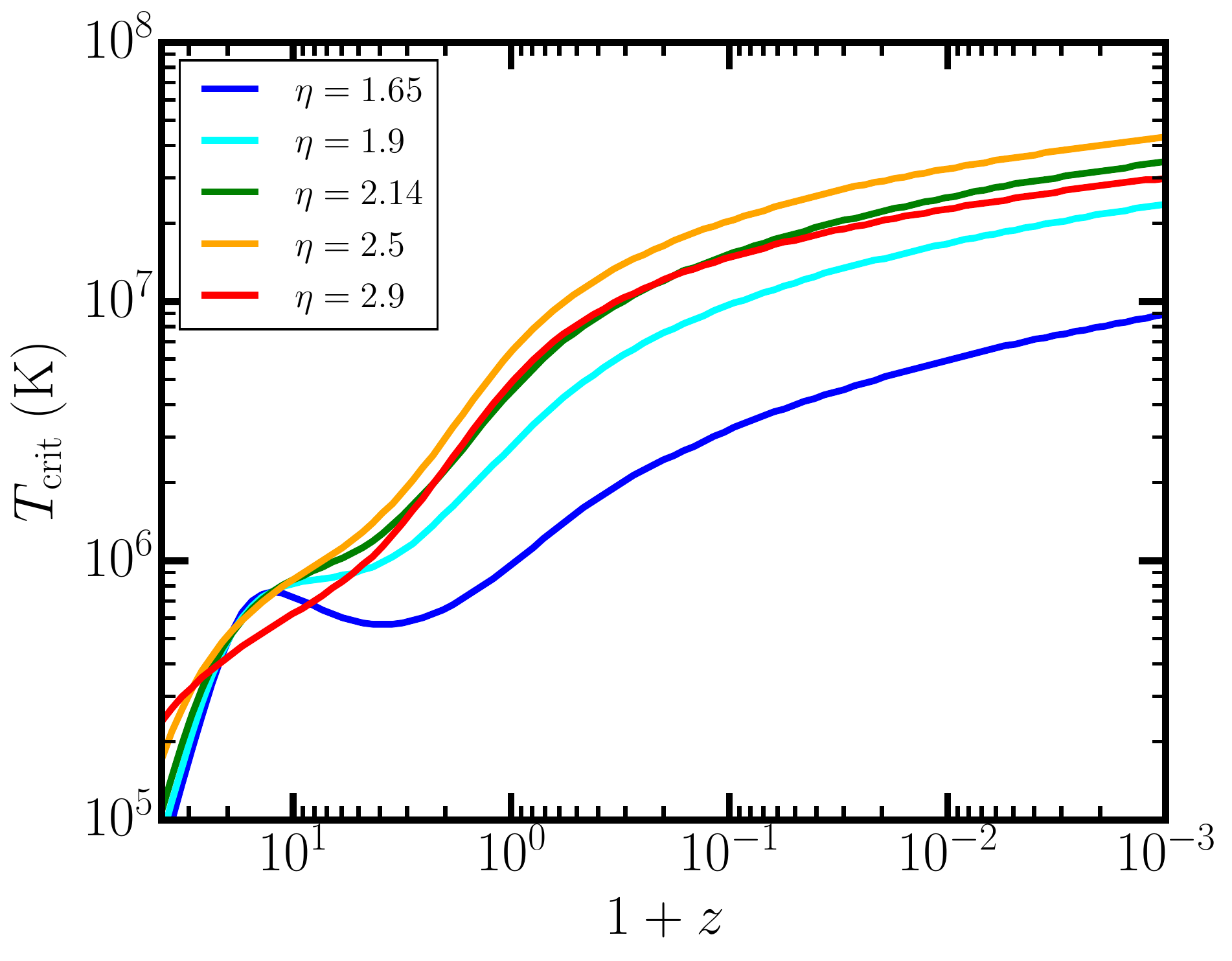}
    \caption{Critical temperature above which the baryon mass fraction of haloes where $1.5< \eta <3$ and $\eta \neq 2$ is equal to the cosmic baryon mass fraction. The solid blue, cyan, green, orange and red lines correspond to $\eta=1.65$, $\eta=1.9$, $\eta=2.14$, $\eta=2.5$ and $\eta=2.9$, respectively.}
    \label{fig:Tcrit}
\end{figure}

Clearly, the physically interesting case corresponds to $1.5<\eta<3$ and $\eta \neq 2$, and we assume this to hold in what follows. We can therefore determine the baryon mass fraction in haloes with virial temperature $T$ and redshift $z$. Indeed, the baryon mass fraction is simply given by the ratio of the gas mass enclosed within a sphere of radius $r_{\rm crit}$, and the total halo mass. The shape of the gas density profile given in equation~\eqref{eq:dens_prof} implies that for $r_{\rm crit}<\min(r_{\rm cool}, \,R)$
\begin{equation}
    f_{\rm b, \, \, halo} (T, \, z) = f_{\rm b} \left( \frac{r_{\rm crit}(T, \, z)}{R} \right)^{3-\eta} \, .
\end{equation}
If we re-cast the right hand side of equation~\eqref{eq:rb} in terms of the virial temperature, the above expression can be written as:
\begin{gather}
\label{eq:fbh_T}
    f_{\rm b, \, \, halo} (T, \, z) =   
        \begin{cases}
         \left(  \frac{T}{T_{\rm crit}(z)}\right) ^{\frac{3-\eta}{2(\eta - 1)}} f_{\rm b} &  \textrm{if} \;\; T<T_{\rm crit}(z) \\
            f_{\rm b} & \rm otherwise
        \end{cases} \, ,
\end{gather}
where $T_{\rm crit} (z)$ is the critical temperature above which gas extends all the way up to the cooling radius, at any given redshift. Thus, in order to find the final expression of $f_{\rm b, \, \, halo} (T, \, z)$, we must determine $T_{\rm crit} (z)$.

This is done by requiring $r_{\rm crit} < \min( r_{\rm cool}, R)$, which in the low-redshift case translates into $r_{\rm crit} < r_{\rm cool}$. Following equations~\eqref{eq:rb} and \eqref{eq:rcool}, adopting the approximation $M_{\rm gas} \approx f_{\rm b,\, halo}(T,\,z) M$ (see \S\,\ref{sec:low-z}-\ref{sec:high-z}), and from the definitions of the virial quantities in equations~\eqref{eq:Rvir}-\eqref{eq:Mvir}, the condition $r_{\rm crit} \leq r_{\rm cool}$ can be re-written as
\begin{equation}
    \label{eq:Tcrit_cond}
    \left( \frac{-\eta^2 + 4\eta -2}{\eta} \frac{GM}{f_{\rm w} \beta x R^2} \right)^{\frac{2\eta-3}{\eta-1}} \leq \frac{(3-\eta) f_{\rm b} \mu X^2 M \Lambda(T)}{6\pi k_{\rm B} T m_{\rm H}^2 R^3} t_{\rm cool}(z),
\end{equation}
where we adopted the definition of the cooling time given by equation~\eqref{eq:tcool_final}.

We solved equation~\eqref{eq:Tcrit_cond} numerically, assuming a primordial cooling function. To avoid abrupt variations of the trend of the critical temperature with redshift between the regime where $r_{\rm cool} < R$ (low redshift) and $r_{\rm cool} > R$ (high redshift) determined by the condition $r_{\rm crit} < \min( r_{\rm cool}, R)$,
we adopted a smooth transition function similar to the one for $s(M,\,z)$ in equation~\eqref{eq:nSFR_smooth}, and we plot the resulting $T_{\rm crit}(z)$ in Figure~\ref{fig:Tcrit}. Because $t_{\rm cool}(z)$ is approximately equal to the age of the universe in the future (see discussion in \S\,\ref{sec:low-z}), it follows that the right hand side of equation~\eqref{eq:Tcrit_cond} becomes arbitrarily large as $z \rightarrow -1$. That is why the critical temperature is monotonically increasing (albeit slowly) at negative redshift in Figure~\ref{fig:Tcrit}. It follows that there is always a time when the critical temperature becomes larger than the virial temperature of a given halo. Therefore, the baryon mass fraction of all haloes will eventually drop below the cosmic baryon fraction, and will keep decreasing thereafter.

To understand this behaviour in more detail, consider a halo with a fixed virial temperature $T$. At late times, $r_{\rm cool}$ grows  monotonically with time (see equation~\eqref{eq:rcool} and Figure~\ref{fig:tcool}). A larger amount of gas is therefore converted into stars; but the stellar winds ejected by the most massive stars will now affect a larger volume, thereby pushing gas away from the halo more effectively and resulting in a lower baryon mass fraction in the halo. Thus, the monotonic increase of $T_{\rm crit}(z)$ tells us that the baryon fraction in all haloes ultimately falls because of gas depletion. It is important to note that this mechanism ignores the contribution of AGN feedback; as discussed in \S~\ref{sec:limitations}, this is likely to accelerate the evacuation of hot gas from the halo, hence the baryon fraction in haloes, and consequently the cosmic star formation history, may be overestimated at late times.

We will study the behaviour of the baryon mass fraction in haloes for different values of $\eta$ in the next section, where we will also compare our predictions with various observations.

\begin{figure}
    \centering
    \includegraphics[width=0.98\columnwidth]{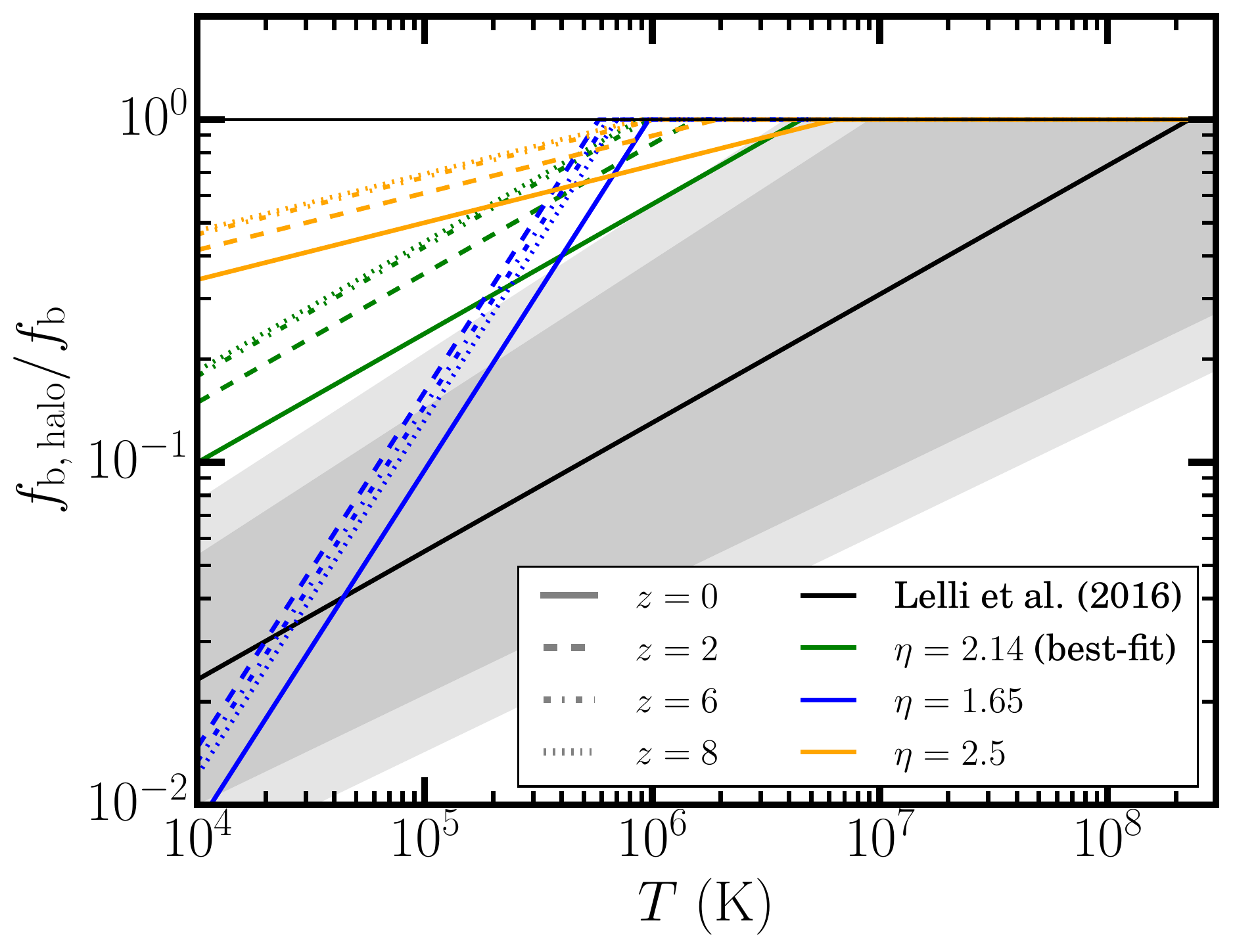}
    \caption{Baryon mass fraction within haloes, in units of the cosmic baryon fraction, as a function of the virial temperature. Solid blue and orange lines refer to $\eta=1.65$ and $2.5$, respectively. The green lines correspond to $\eta=15/7\approx 2.14$, which provides the best match to the slope of the bTFR measured by \protect\cite{Lelli_2016}, shown with the solid black line. The dark grey shaded area around the black line represents the region where the best-fit parameters to the observed bTFR have been varied within $1\sigma$. The light grey shaded area shows the extra scatter on the $f_{\rm b,\, halo} - T$ relationship inferred from \protect\cite{Lelli_2016} observations due to the uncertainty on the conversion factor between the  measured circular velocity in the flat region of the rotation curve ($V_{\rm f}$) and the virial velocity $V_{200}$ (see main text for details). For each value of $\eta$ considered, we plot the predictions for $z=0$, $z=2$, $z=6$ and $z=8$, which are indicated with solid, dashed, dot-dashed and dotted lines. The horizontal black line serves only as a guide line, and corresponds to $f_{\rm b, \, halo} = f_{\rm b}$. We plotted $f_{\rm b,\, halo} (T, \, z)$ with the definition given by equation~\eqref{eq:fbh_T} to clearly show the position of the knee of the bTFR. However, in all our calculations we then smooth between the $T<T_{\rm crit}(z)$ and $T>T_{\rm crit}(z)$ regimes, similarly to the treatment of $s(M, \, z)$ in equation~\eqref{eq:nSFR_smooth}.
    }
    \label{fig:fbh_T}
\end{figure}

\subsection{Comparison with observations}

We plot the baryon mass fraction in haloes given by equation~\eqref{eq:fbh_T} in units of the cosmic baryon fraction in Figure~\ref{fig:fbh_T}. 
For every value of $\eta$ in this plot, the critical temperature moves towards lower values at higher redshifts, consistent with Figure~\ref{fig:Tcrit}. We notice that the evolution with redshift is stronger for larger values of $\eta$, again in concordance with Figure~\ref{fig:Tcrit}. The slope of the relationship for $T<T_{\rm crit} (z)$ is independent of redshift, but the evolution does depend quite strongly on $\eta$, with smaller values of $\eta$ corresponding to steeper slopes, as dictated by equation~\eqref{eq:fbh_T}.

We can ask ourselves which values of $\eta$ provide the best match to the relationship between $f_{\rm b, \, halo}$ and the virial temperature inferred from the observations of the bTFR by \cite{Lelli_2016}. Originally, \cite{Lelli_2016} provided a fit to the correlation between the baryon mass in haloes and the rotation velocity in the flat region of the rotation curve ($V_{\rm f}$) that they measured in a sample of galaxies:
\begin{equation}
\label{eq:Lelli}
   \log_{10} \left( \frac{M_{\rm b}}{\rm M_{\odot}} \right) = \log_{10} Q + p \log_{10} \left(\frac{V_{\rm f}}{\rm km \, s^{-1}} \right) \, ,
\end{equation}
finding $p=(3.75 \pm 0.11)$ and $\log_{10} Q =(2.18\pm 0.23)$ for their accurate-distance sample (for details, see \citejap{Lelli_2016}). The question for the present paper is then how to relate $V_{\rm f}$ to the virial temperature. \cite{Lelli_2016} argued that $V_{\rm f}$ is a good proxy for $V_{200}$, i.e. the virial velocity associated with a halo with virial mass and virial radius given by our equations~\eqref{eq:Rvir}-\eqref{eq:Mvir} with $\Delta =200$. Thus, they suggested that $V_{\rm f} = f_V V_{200}$, with $f_V \approx 1$. However, they also point out that the bTFR has a scatter of at least $\sim 0.15 \, \rm dex$ in $M_{\rm b}$, based on the results of N-body simulations \citep{Moster_2013, Dutton_2014} and semi-analytic models \citep{Dutton_2012, Zu_2015}. This drives a scatter in $f_V$, meaning that values $0.9 \lesssim f_V \lesssim 1.1$ are all sensible, but this range may actually be wider. In fact, in an earlier similar work \cite{McGaugh_2010} argued that even $f_V \approx 1.3$ would be reasonable. Using the parametrisation $V_{\rm f}= f_V V_{200}$, and dividing both sides of equation~\eqref{eq:Lelli} by the virial mass given by equation~\eqref{eq:Mvir} with $\Delta=200$, we have:
\begin{equation}
\label{eq:Lelli_T}
    f_{\rm b,\, halo} = 10 \, Q \frac{G H_0 \mathrm{M}_{\odot}}{\rm km^3 \, s^{-3}} f_{\rm V}^p \left(\frac{2 k_{\rm b} T}{\mu\, \rm km^2 \, s^{-2}} \right)^{\frac{p-3}{2}} \, ,
\end{equation}
where we used the definition of virial temperature in equation~\eqref{eq:Vvir} to express the virial mass and velocity in terms of $T$.

We can now plot the best fit to the bTFR measured by \cite{Lelli_2016} in the $f_{\rm b, \, halo} - T$ diagram in Figure~\ref{fig:fbh_T}: this corresponds to the black solid line. The dark grey shaded area encompasses the region between the relationships defined by equation~\eqref{eq:Lelli_T} with $(p=3.86, \,\log_{10} Q= 2.41)$ and $(p=3.64, \,\log_{10} Q= 1.95)$, i.e., where the parameters of the observed bTFR have been varied within $1\sigma$ with respect to the best-fit values. The light grey shaded area shows the extra scatter on the relationship once the uncertainty on $f_V$ is taken into account: specifically, we have considered $0.9 \leq f_V\leq 1.1$.

The best match with the observed slope for $T<T_{\rm crit}$ is achieved for $\eta=15/7 \approx 2.14$. The $f_{\rm b\, halo}-T$ relationship given by equation~\eqref{eq:fbh_T} for this value of $\eta$ is also plotted in Figure~\ref{fig:fbh_T}  (green lines). Even though we determined $\eta=15/7$ to be the value that best matches the slope of the relationship, we notice that the critical temperature at $z=0$ is off by $\sim 1.5\, \rm dex$ with respect to the the value extrapolated from the bTFR observations. However, the large uncertainties on the normalisation of the bTFR should be taken into account here. To assess how reasonable our predicted critical temperature is, we should really consider a data set that covers the knee of the bTFR, avoiding the need for extrapolation. With this in mind, we note that our critical temperature corresponds to a circular velocity of $353\, \rm km \, s^{-1}$, agreeing at the percent level with the observed knee of the bTFR obtained by \cite{McGaugh_2010} from a compilation of observations spanning three orders of magnitude in the circular velocity of galaxies and clusters ($20 \,  \mathrm{km} \, \mathrm{s}^{-1}\lesssim V_{\rm c} \lesssim 2000 \, \rm km \, s^{-1}$). This is a reassuring and non trivial result, since matching $\eta$ to the slope observed by \cite{Lelli_2016} does not in principle guarantee matching the knee of the bTFR. Our model thus provides a pleasingly simple consistent physical explanation for the bTFR.

The $f_{\rm b, \, halo}-T$ relationship that we obtain for $\eta=15/7$ also succeeds in explaining a completely different kind of observation, namely the Kennicutt-Schmidt relationship \citep{Kennicutt_1998}. The power-law index of the relation between the surface SFR density and gas surface density was found to be $N=1.4\pm0.15$ by \cite{Kennicutt_1998}, and attempts were made to give a theoretical justification for values around $1.4-1.5$ \citep[e.g.][]{Krumholz_2009, Renaud_2012, Kraljic_2014}. Our simple model for the baryon fraction in haloes, combined with our extended Hernquist-Springel formalism for star formation, succeeds in providing such an explanation. From equation~\eqref{eq:SFR}, we have that at any fixed (low) redshift \smash{$\Sigma_{\rm SFR} \propto \Sigma_{\rm gas}^{3/\eta}$}. For $\eta=15/7$, we obtain exactly \smash{$\Sigma_{\rm SFR} \propto \Sigma_{\rm gas}^{1.4}$}. It is noteworthy that by adjusting $\eta$ to reproduce the slope of the \cite{Lelli_2016} observations we also obtain the correct Kennicutt-Schmidt relationship. Nevertheless, other observations showed that the index can vary in the range $(1.1, 1.7)$, depending on the redshift and type of galaxies in the observed sample \citep[e.g.][]{Bouche_2007, Bothwell_2010, Genzel_2010, Azeez_2016}. Within our formalism, this range would correspond to slopes of the gas density profile $1.8 \lesssim \eta \lesssim 2.7$. These constraints on $\eta$ include the range of $\eta$ that is consistent with the \cite{Lelli_2016} observations within $3\sigma$, i.e. $1.92 \lesssim \eta \lesssim 2.41$.

\begin{figure}
    \centering
    \includegraphics[width=0.98\columnwidth]{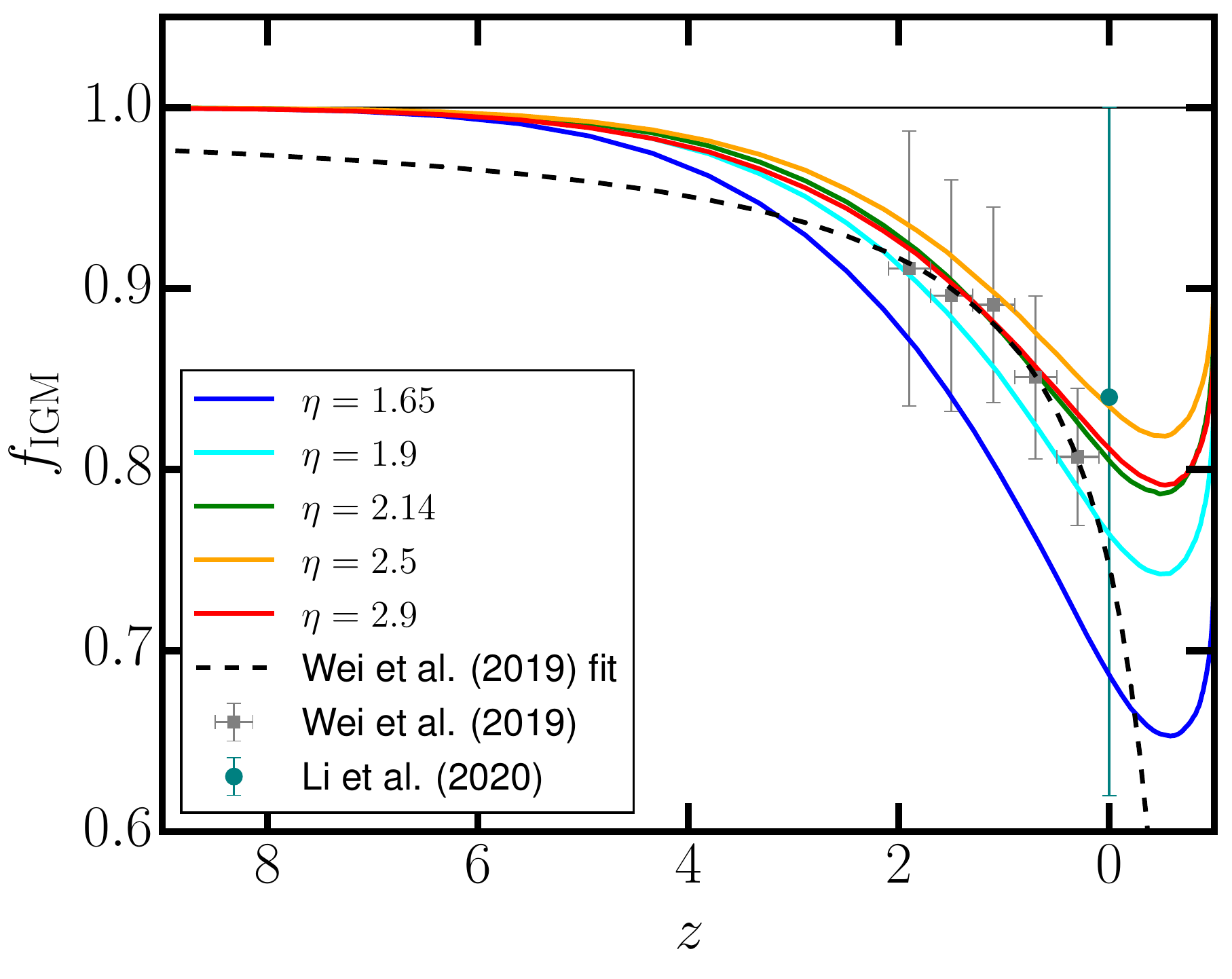}
    \caption{Baryon mass fraction in the IGM predicted by our model for different values of the slope of the gas density profile within haloes $\eta$, following the same colour coding as in Figure~\ref{fig:Tcrit}. The black solid line shows the fit to the baryon mass fraction in the IGM obtained from mock FRB dispersion measures (grey squares), provided by \protect\cite{Wei_2019}. The horizontal error bars mark the bin widths in the mocks. The teal circle represents the $f_{\rm IGM}$ at redshift $z=0$ measured by \protect\cite{Li_2020}.  The horizontal thin solid line marks $f_{\rm IGM}=1$, to guide the eye. Given the constraints on $f_{\rm IGM}$, our model gives reasonable predictions for all values of $\eta$.}
    \label{fig:fIGM}
\end{figure}

Although our model yields reasonable results for the baryon fraction of haloes, we should check whether this is still the case when we consider the the baryon mass fraction in the intergalactic medium (IGM), which  we denote as $f_{\rm IGM}(z)$. While it is well known that $\sim 90\%$ of the baryons are locked in the IGM at $z\gtrsim 1.5$ \citep{Meiksin_review, McQuinn_2016}, the value of $f_{\rm IGM}(z)$ at lower redshift is not yet precisely known, and as such it is the subject of ongoing research \citep[e.g.][]{Li_2019, Li_2020, Qiang_2020}. We show the recent measurement of $f_{\rm IGM}(z)$ from observations of fast radio bursts (FRBs) at $z\sim 0$ by \cite{Li_2020} with the teal circle in Figure~\ref{fig:fIGM}. We also show the fit (dashed black line) by \cite{Wei_2019} to the $f_{\rm IGM}(z)$ values obtained from mock FRB dispersion measures at $z<2.1$ (grey squares). The horizontal error bars of the grey points represent the bin widths utilised in the mock data.

We compare compare the results from these works with the predictions of our model, where we computed $f_{\rm IGM} (z)$:
\begin{equation}
    f_{\rm IGM}(z) = 1 - \int f_{\rm b,\, halo}(T(M,\,z), z) \frac{dF}{d\ln M} d\ln M \, .
\end{equation}
The simplifying assumption in the above equation is that all gas in the universe can be found either in the IGM or in haloes. We integrated from the minimum mass of star-forming haloes, as explained in \S\,\ref{sec:CSFRD}.

Our predictions for $f_{\rm IGM }$ relative to different values of $\eta$ are plotted in Figure~\ref{fig:fIGM}, with the same colour coding as in Figure~\ref{fig:Tcrit}. For all values of $\eta$, $f_{\rm IGM}$ is above $\sim 90\%$ at high redshift ($z\gtrsim 1.5$), in good accord with early observations and simulations (see the review by \citejap{Meiksin_review}). This behaviour reflects the fact that most baryons are yet to collapse in haloes with $M>M_4$ at high redshift. At lower redshift, more and more baryons contribute to halo accretion, hence lowering $f_{\rm IGM}$. Depending on the value of $\eta$, at $z=0$ the fraction of baryons in haloes lies between $\sim0.2$ and $0.3$. All models agree with the \cite{Li_2020} data point, given the size of the error bars. The results are also in broad agreement with the baryon mass fraction in low-redshift haloes found by \cite{Shull_2012}. We notice that values $\eta \geq 1.9$ are preferred by \cite{Wei_2019} mock data at $0<z<2$, but actually all values of $\eta$ are consistent with such mocks, given the size of the error bars. The only exception might be $\eta=1.65$, which is somewhat in tension with the lowest-redshift mocks. Future observations will provide a larger number of FRBs, which will better constrain $f_{\rm IGM}$, perhaps allowing us to exclude certain values of $\eta$ from our model.

We notice that in the far future $f_{\rm IGM}$ is predicted to increase again. The reason is that the critical temperature keeps increasing with cosmic time (see Figure~\ref{fig:Tcrit}), so that an ever larger fraction of haloes will have their baryon mass fraction suppressed, as dictated by equation~\eqref{eq:fbh_T}. Indeed, values of $\eta$ that yield a lower critical temperature (e.g., $\eta=1.65$) are associated with lower $f_{\rm IGM}$ as $z\rightarrow -1$. The physical interpretation of this long-term behaviour of the gas mass fraction in the IGM is that in the far future stellar winds will have overtaken star formation via gas cooling within haloes, eventually depleting haloes with gas and quenching further star formation. 

We caution that in Figure~\ref{fig:Tcrit} we simply extrapolated the \cite{Wei_2019} fit to high redshift, even though the empirical fit was derived from mock data at redshift $0<z<2$. While at high redshift it is reasonable to have $f_{\rm IGM}>0.9$, the fit cannot be necessarily trusted for $z<0$.

To sum up, we believe that our analytic model for the baryon mass fraction in haloes effectively captures the main aspects of the missing baryon problem. It succeeds in reproducing observations of the bTFR \citep{Lelli_2016, McGaugh_2010}, the Kennicutt-Schmidt relationship \citep{Kennicutt_1998}, and the baryon fraction of the IGM both at low \citep{Shull_2012, Li_2020} and high redshift \citep[see review by][]{Meiksin_review}. Given the simplicity of the model, this overall agreement with several diverse observations is pleasing.

%%%%%%%%%%%%%%%%%%%%%%%%%%%%%%%%%%%%%%%%%%%%%%%%%%

% Don't change these lines
\bsp	% typesetting comment
\label{lastpage}
\end{document}